\documentclass[11pt]{article}
\newtheorem{proposition}{Proposition}

\usepackage[toc,page]{appendix}
\newenvironment{proof}{\noindent{\bf Proof:}}{\hfill\fbox{}\vspace*{1mm}}
\usepackage[usenames]{color}
\usepackage{color}
\usepackage{amssymb}
\usepackage{graphicx}
\usepackage{subfigure}
\usepackage{float}
\usepackage{booktabs}

\RequirePackage[normalem]{ulem} 
\providecommand{\DIFdeltex}[1]{{\protect\color{red}\sout{#1}}}                      
\usepackage{xcolor}
\usepackage{float}
\usepackage{epstopdf}
\usepackage{changebar}
\newif\ifdiff
\difftrue
\ifdiff
  \newcommand{\del}[1]{\DIFdeltex{#1}}

\else
  \newcommand{\del}[1]{}

\fi

\begin{document}
\title{\bf Interacting Default Intensity with Hidden Markov Process}
\author{
Feng-Hui Yu
\thanks{ Advanced Modeling and Applied Computing Laboratory,
Department of Mathematics,
The University of Hong Kong, Pokfulam Road, Hong Kong.
E-mail: maggie.yufenghui@gmail.com.}
\and Wai-Ki Ching
\thanks{Corresponding author.
Advanced Modeling and Applied Computing Laboratory,
Department of Mathematics,
The University of Hong Kong, Pokfulam Road, Hong Kong.
E-mail: wching@hku.hk. }
\and Jia-Wen Gu
\thanks{Department of Mathematical Science,
University of Copenhagen,
Denmark.
E-mail: jwgu.hku@gmail.com.}
\and Tak-Kuen Siu
\thanks{ Department of Applied Finance and Actuarial Studies,
Faculty of Business and Economics,
Macquarie University, Sydney, NSW 2109, Australia.
Email: Ken.Siu@mq.edu.au;ktksiu2005@gmail.com}
}

\maketitle

\begin{abstract}
In this paper we consider a reduced-form intensity-based credit risk model with a hidden Markov state process.
A filtering method is proposed for extracting the underlying state given the observation processes.
The method may be applied to a wide range of problems.
Based on this model, we derive the joint distribution of multiple default times without imposing stringent assumptions on the form of default intensities.
Closed-form formulas for the distribution of default times are obtained which are then applied to solve a number of practical problems such as hedging and pricing credit derivatives.
The method and numerical algorithms presented may be applicable to various forms of default intensities.
\end{abstract}


\section{Introduction}

Modeling credit risk has long been a critical issue in credit risk management. Attention has been given
 to it especially since the global financial crisis in 2008.
Credit risk modeling has a lot of applications, for example,
pricing and hedging the credit derivatives,
as well as the management of credit portfolios.
Models adopted in the finance industry may be grouped into two major categories:
structural firm value models and reduced-form intensity-based models.
For the first class of models, it was pioneered by Black and Scholes (1973) 
and Merton (1974).
The key idea of the structural firm's value model is to model the default of a firm by using its asset value,
where the asset value is governed by a geometric Brownian motion.
When the asset value falls below a certain prescribed level,
the default of the firm is triggered.
For the second kind of model, it was pioneered by Jarrow and Turnbull (1995) 
and Madan and Unal (1998).
The main idea of reduced-form intensity-based models is to consider the defaults as exogenous processes
and describe their occurrences  with Poisson processes and their variants.

The interacting intensity-based default models are widely adopted to model
the portfolio credit risk and defaults.
Since we focus on contagion models in this paper as in, for example, Giesecke (2008),
we differentiate intensity-based credit risk models into top-down models and bottom-up models.
The top-down models focus on modeling the default times at the portfolio level without reference to the intensities of individual entities.
Based on this, one can also recover the individual entity's intensity with some method like random thinning, etc.
Some works related to this class of models include
Davis and Lo (2001),
Giesecke, Goldberg and Ding (2005),
Brigo, Pallavicini and Torresetti (2006),
Longstaff and Rajan (2008)
and Cont and Minca (2011), etc.
While the bottom-up model focuses on modeling the default intensities of individual reference entities and their aggregation to form a portfolio default intensity.
Some works related to this class of models include
Duffie and Garleanu (2001),
Jarrow and Yu (2001),
Sch\"onbucher and Schubert (2001),
Giesecke and Goldberg (2004),
Duffle et al. (2006)
and Yu (2007), etc.
The differences between these two classes of models are the form of individual
entity's default intensities and the way the portfolio aggregation is formed.
In this paper we shall focus on a bottom-up model.

Based on the model developed by Lando (1998), Yu (2007)
extended the model and  applied the extended model multiple defaults and their correlation.
In addition, Yu adopted the total hazard construction method proposed
by Norros (1986) and Shaked and Shathanthikumar (1987)
to simulate the distribution of default times which have interacting intensities.
Zheng and Jiang (2009) then adopted this method and derived closed-form formulas for the multiple default distributions under their contagion model.
Gu et al. (2013) introduced a recursive method to calculate the distribution of ordered default times, and Gu et al. (2014) further proposed a hidden Markov reduced-form model with a specific form of default intensities.

In this paper we develop a generalized reduced-form intensity-based credit model with hidden Markov process.
The model is applicable to a wide class of default intensities with various forms of dependent constructions.
For the hidden Markov process, we also discuss a flexible method to extract the hidden state process given the observations processes, which may hopefully have applications in diverse fields.
Then using the total hazard construction method by Yu (2007), we derive closed-form formulas for the joint default distribution. When the intensities are homogeneous, analytic algorithm for the calculation
of the joint distribution of ordered default times is provided. The explicit formula may enhance the computational efficiency in applications, for instance, pricing of credit derivatives. We remark that the results in Gu  et al. (2014) is a special case of the method discussed here. In addition, we extend the total hazard construction method to the cases with hidden process to simulate the joint distribution of default times.
We remark that hidden Markov models have been employed in studying credit risk,
see for instance, Frey and Runggaldier (2010, 2011), Frey and Schmidt (2011), Elliott and Siu (2013) and Elliott et al. (2014).

The rest of this paper is structured as follows.
Section \ref{Model Setup} gives a snapshot of the interacting intensity-based default model with hidden Markov process.
Section \ref{Extraction of Hidden Process with Observable Processes} presents the method for extracting the hidden state process from the observation processes.
Section \ref{Default Distributions} derives the closed-form expression for the joint default distribution based on the total hazard construction method, and also presents an analytic formula for the distribution of ordered default times. Besides, the extended total hazard construction method under a hidden Markov process to obtain the joint distribution of default times is also presented. Section \ref{Numerical Method} provides numerical methods for some situation in Section \ref{Extraction of Hidden Process with Observable Processes} which may be used in both Sections \ref{Extraction of Hidden Process with Observable Processes} and \ref{Default Distributions},
and error analysis is also discussed.
Section \ref{Numerical Experiments} illustrates an application of the proposed method in pricing credit derivatives.
Finally, Section \ref{Concluding Remarks} concludes the paper.

\section{Model Setup}\label{Model Setup}

Let $(\Omega, \mathcal{F}, P)$ be a complete probability space
where $P$ is a risk-neutral probability measure, which is assumed to exist.
Suppose there are  $K$ interacting entities, and we let $N_i(t):=1_{\{\tau_i \leq t\}}$,
where $\tau_i$ is a stopping time, representing the default time of credit name $i$, for each $i = 1, 2, \cdots, K$.
Suppose we have an underlying state process $(X_t)_{t\geq 0}$
describing the dynamics of the economic condition.
Let $\mathcal{F}_t^X:=\sigma(X_s, 0\leq s\leq t)\vee\mathcal{N}$
where $\mathcal{N}$ represents all the null subsets of $\Omega$ in ${\cal F}$ and ${\cal C}_1 \vee {\cal C}_2$ is the minimal $\sigma$-algebra containing both the $\sigma$-algebras ${\cal C}_1$ and ${\cal C}_2$.
We also let
$\mathcal{H}_t:=\sigma(X_t) \vee \mathcal{F}^N_t$ where
$$
\mathcal{F}^N_t=\mathcal{F}^1_t\vee\mathcal{F}^2_t\vee \ldots \vee \mathcal{F}^K_t
\quad {\rm and} \quad
\mathcal{F}^i_t:=\sigma(1_{\{\tau_i \leq s\}}, 0 \leq s \leq t)\vee\mathcal{N}.
$$
We assume that for each $i = 1, 2, \ldots, K$,
${N_i}(t)$ possesses a nonnegative, $\{ \mathcal{H}_t \}_{t \geq 0}$-adapted,
intensity process $\lambda_i$ satisfying
\begin{equation}
E\left( \int_0^t \lambda_i(s) ds \right) < \infty, \quad  t\geq 0,
\end{equation}
such that the compensated process
\begin{equation}
M_i(t) := N_i(t)-\int_0^{t \wedge \tau_i} \lambda_i(s) ds \ , \quad t \geq 0,
\end{equation}
is an $(\{\mathcal{H}_t\}_{t \geq 0}, P)$-martingale.
Note that after the default time $\tau_i$, $N_i (t)$ will stay at the value one, so
there is no need to compensate for $N_i (t)$ after time $\tau_i$, see, for example, Elliott et al. (2000).

For all the market participants, we assume that they cannot observe the underlying
process $(X_t)_{t \geq 0}$ directly.
Instead, they observe the process $(Y_t)_{t\geq 0}$,
revealing the delayed and noisy information of $(X_t)_{t\geq 0}$,
and also observe the default process $(N_t^i)_{t\geq0}$.
Hence, the common information set available to the market participants at time $t$ is
$\mathcal{F}_t:=\mathcal{F}^Y_t \vee \mathcal{F}^N_t$
where $\mathcal{F}_t^Y:=\sigma(Y_s, 0\leq s\leq t)\vee\mathcal{N}$.
We further assume that $(X_t)_{t\geq 0}$
is an ``exogenous'' process to  $(N_t^i)_{t\geq0}, i=1,2,\ldots, K$, i.e., For any $t$, the $\sigma$-fields
$\mathcal{F}_{\infty}^X$
and ${\mathcal{F}}_t^N$
are conditionally independent
given ${\mathcal{F}}_t^X$ and
$P(\tau_i \neq \tau_j)=1, i \neq j$.

To simplify our discussion, throughout the paper, we suppose that $(X_t)_{t \geq 0}$
is a two-state Markov chain taking a value in $\{x_0, x_1\}$.
We assume the transition rates of the chain for
``$x_0\rightarrow x_1$'' and ``$x_1 \rightarrow x_0$'' are
$\theta_{0}$  and $\theta_{1}$, respectively.
The observable process $(Y_t)_{t\geq 0}$ is again a two-state Markov chain
taking value in $\{y_0, y_1\}$, with transition rates depending on $X_t$,
i.e., $\eta_{0}(X_t)(y_0 \rightarrow y_1)$
and $\eta_{1}(X_t)(y_1\rightarrow y_0)$,
where $\eta_{0}$ and $\eta_{1}$ are real-valued functions.
At time $0$, we suppose that $X_0$ is in state $x_0$ and $Y_0$ is in state $y_0$.
The methods introduced later in our paper may still be
applicable when the Markov chains $X$ and $Y$ have more than two states though more complicated notation may involve.

\section{Extraction of Hidden State Process with Observable Processes}\label{Extraction of Hidden Process with Observable Processes}

To specify the form of the intensities, we give the following notations.
Suppose that at time $t$, $N_t^D$ defaults have already occurred at
$t_1,t_2,\ldots,t_{N_t^D}$ such that
$$
0=t_0<t_1< \cdots < t_{N_t^D}\le t.
$$
Then we denote $T_{N_t^D}=(t_1,\cdots,t_{N_t^D})$ the ordered $N_t^D$ default times
and $I_{N_t^D}=(j_1,\cdots,j_{N_t^D})$ the corresponding $N_t^D$ defaulters,
and the $m$th $(1\le m\le K)$ defaulted obligor is $j_m$.
We assume that $i>N_t^D$ and $t<\tau^i$, where $\tau^i$ is the obligor $i$'s default time. Each process $\lambda_i$ ($i=1, \ldots, K$), is $\{ \mathcal{H}_t \}_{t \geq 0}$-predictable, that is to say $\lambda_i (t)$ is known given information about the chain $X$ and all the default processes prior to time $t$.
Then the intensity of $\tau^i$ may be written as
{$\lambda_t^i=\lambda_i(t|I_{N_t^D}, T_{N_t^D}, X_t)$}
where $X_t$ is the state of chain $X$ at time $t$.
Note that $(I_{N_t^D}, T_{N_t^D}, X_t)\in \mathcal{H}_t$.

Since the path of $X$ is unobservable,
while the path of $Y$ and $N_i$, ($i=1,\ldots, K$) are observable,
we can use the relationship between $X$, $Y$ and $N_i$, ($i=1,\ldots, K$)
to find the probability law of $X$.
We apply the recursive method proposed in Gu et al. \cite{Gu2}
to calculate the conditional probability $P(X_t=x_i|\mathcal{F}_t)$, ($i=0, 1, t\ge 0$). Before discussing the method, we need to find the expressions for all
the unknown items in the recursive formulas.
In the process of finding the expressions,
we also present moment generating function method to achieve our goal.

\subsection{Some Preliminaries} \label{Preliminary}

Let $\bar{T}_{i,k,j}(s_0, \Delta s)$ be the union of subintervals of time of
the chain $X$ in state $x_k$ in the time interval
$[s_0, s_0+\Delta s]$ given the chain starts from $X_{s_{0}} =x_i$
and ends at $X_{s_{0}+\Delta s}=x_j$.
For each $i,j= 0, 1$, we let
$$
\bar{T}_{i,j}(s_0, t)=(\bar{T}_{i,0,j}(s_0, t), \bar{T}_{i,1,j}(s_0, t))^T
\quad {\rm and} \quad
{\bf u}(\bar{t})=(u_{0}(\bar{t}),u_{1}(\bar{t}))^T
$$
where $\bar{t}\in [s_0, s_0+t]$.
Note that $\bar{T}_{i,1,j}(s_0, t)=[s_0, s_0+t]\backslash \bar{T}_{i,0,j}(s_0, t)$.
Since jumps in chain $Y$ and defaults are Poisson processes,
using the concept of moment generating function, we define
$$
\bar{\Psi}_{ij}(s_0, {\bf u},t)
=E \left[\exp\left\{\int_{\bar{T}_{i,0,j}(s_0, t)}u_{0}(\bar{t})d\bar{t}+\int_{\bar{T}_{i,1,j}(s_0, t)}u_{1}(\bar{t})d\bar{t}\right\}\right].
$$
Note that ${\bf u}(\bar{t})$ is an arbitrary integrable function.
This means, in this case, we can adopt this moment generating function.
For instance, ${\bf u}(\bar{t})$ can be the transition rates of jumps in chain $Y$
or the default rates which are the default intensities accumulated by all the entities by time $\bar{t}$ before default.

\begin{proposition}
Let $\bar{\Phi}_{ij}(s_0, {\bf u}, t)=P_{ij}(t)\bar{\Psi}_{ij}(s_0, {\bf u}, t)$, where $P_{ij}(t)$ is the probability that a process in state $x_i$ will be in state $x_j$ after a time of $t$, and $i, j=0, 1$. Then
\begin{equation}\label{f5}
\left\{
\begin{array}{l}
\displaystyle \bar{\Phi}_{ij}(s_0, {\bf u}, t)=\theta_i \int_0^t \exp \left(\int_{s_0}^{s_0+t-s}(u_i(\bar{t})-\theta_i)d\bar{t}\right)
\bar{\Phi}_{jj}(s_0+t-s, {\bf u}, s) ds\\
\displaystyle \bar{\Phi}_{ii}(s_0, {\bf u}, t)=\theta_i \int_0^t \exp\left(\int_{s_0}^{s_0+t-s}(u_i(\bar{t})-\theta_i)d\bar{t}\right)
\bar{\Phi}_{ji}(s_0+t-s, {\bf u}, s) ds+ \exp\left(\int_{s_0}^{s_0+t}(u_i(\bar{t})-\theta_i)d\bar{t}\right)
\end{array}
\right.
\end{equation}
where $i,j=0,1$.
\end{proposition}

\begin{proof}
$$
\begin{array}{lll}
&\bar{\Psi}_{ij}(s_0, {\bf u},t)\\
=&\displaystyle E\left[\exp\left(\int_{\bar{T}_{i,0,j}(s_0, t)}u_{0}(\bar{t})d\bar{t}+\int_{\bar{T}_{i,1,j}(s_0, t)}u_{1}(\bar{t})d\bar{t}\right)\right]\\
=&\displaystyle \frac{\theta_i}{P_{ij}(t)}\int_0^t e^{-\theta_is}\cdot e^{\int_{s_0}^{s_0+s}u_i(\bar{t}) d\bar{t}} P_{jj}(t-s)E\left[\exp\left(\int_{\bar{T}_{j,0,j}(s_0+s, t-s)}u_0(\bar{t})d\bar{t}+\int_{\bar{T}_{j,1,j}(s_0+s, t-s)}u_1(\bar{t})d\bar{t}\right)\right] ds\\
=&\displaystyle \frac{\theta_i}{P_{ij}(t)}\int_0^t \exp\left(\int_{s_0}^{s_0+s}(u_i(\bar{t})-\theta_i)d\bar{t}\right)P_{jj}(t-s) {\bar{\Psi}}_{jj}(s_0+s, {\bf u}, t-s) ds\\
=&\displaystyle \frac{\theta_i}{P_{ij}(t)}\int_0^t \exp\left(\int_{s_0}^{s_0+t-s}(u_i(\bar{t})-\theta_i)d\bar{t}\right)P_{jj}(s) {\bar{\Psi}}_{jj}(s_0+t-s, {\bf u}, s) ds.
\end{array}
$$
We also have
$$
\bar{\Psi}_{ii}(s_0, {\bf u},t)=\frac{\theta_i}{P_{ii}(t)}\int_{0}^{t}e^{\int_{s_0}^{s_0+t-s}(u_i(\bar{t})-\theta_i)d\bar{t}}P_{ji}(s)\bar{\Psi}_{ji}(s_0+t-s, {\bf u},s)ds+\frac{e^{\int_{s_0}^{s_0+t}(u_i(\bar{t})-\theta_i)d\bar{t}}}{P_{ii}(t)}.
$$
Replace $\bar{\Psi}_{ij}(s_0, {\bf u}, t)$ by $\frac{\bar{\Phi}_{ij}(s_0, {\bf u}, t)}{P_{ij}(t)}$,
we can then get the system of equations in the proposition.
\end{proof}

We find that when the expression of ${\bf u}(\bar{t})$ satisfies some ``good'' property,
Eq. (\ref{f5}) in the above proposition has a unique solution.
{The property is that ${\bf u}(\bar{t})$ does not have any direct relationship with time $\bar{t}$ even though it may have implied relationship with $\bar{t}$.
This means ${\bf u}(\bar{t})$ can be written as ${\bf u}$.}
Then, not only the problem of solving Eq. (\ref{f5}) can be simplified,
but some related definitions can also be simplified as well.
Similar as before, let {$T_{i,k,j}(\Delta s)$} be the occupation time of
the chain $X$ in state $x_k$ in the time interval
$[s, s+\Delta s]$ given the chain starting from $X_s =x_i$
and ending at $X_{s+\Delta s}=x_j$. For each $i,j= 0, 1$, we let
$$
T_{i,j}(t)=(T_{i,0,j}(t), T_{i,1,j}(t))^T
\quad {\rm and} \quad
{\bf u}=(u_{0},u_{1})^T \in \mathbb{R}^2.
$$
The moment generating function of $T_{i,j}(t)$ is given by
$$
\Psi_{ij}({\bf u},t)=E(\exp\{{\bf u}^{T} T_{i,j}(t)\}).
$$
Apply the same method to $\Psi_{ij}({\bf u},t)$ as we have done to $\bar{\Psi}_{ij}({\bf u},t)$,
and let
$$
\Phi_{ij}({\bf u},t)=\Psi_{ij}({\bf u},t)\cdot P_{ij}(t).
$$
We can also get the equivalent Eq. (\ref{f5}) for $\Phi_{ij}({\bf u},t)$, $i.e.$,
replacing $\bar{\Phi}_{ij}({\bf u},t)$ with $\Phi_{ij}({\bf u},t)$, $(u_i(\bar{t})-\theta_i)$ with $u_i-\theta_i$ in Eq. (\ref{f5}).
Then to solve the equivalent equation, it suffices to solve a linear
system of O.D.E.s (c.f. Gu et al. \cite{Gu2}):
$$
\frac{\partial \Phi({\bf u},t)}{\partial t} =A \Phi({\bf u},t),
$$
where
$$
\Phi({\bf u},t)=\left[
\begin{array}{cc}
\Phi_{11}({\bf u},t) & \Phi_{12}({\bf u},t)\\
\Phi_{21}({\bf u},t) & \Phi_{22}({\bf u},t)
\end{array}
\right]
\quad {\rm and} \quad
A=\left[
\begin{array}{cc}
u_{0}-\theta_0 & \theta_0\\
\theta_1 & u_{1}-\theta_1
\end{array}
\right].
$$
This linear system of O.D.E.s is known as the fundamental matrix equation in the literature. Then it is well-known that the equation has a unique solution which is called the fundamental matrix solution with the initial condition
$\Phi_{ij}({\bf u}, 0)=1, i,j=0, 1$ as
$$
\Phi({\bf u},t)=e^{At} {\bf 1} \cdot {\bf 1}^T
$$
where
${\bf 1}$ is the two-dimensional column vector
with all entries being equal to $1$.
Hence we can get the solution for $\Psi_{ij}({\bf u},t)$ by
$$
\Psi_{ij}({\bf u},t)=\frac{\Phi_{ij}({\bf u},t)}{P_{ij}(t)}.
$$

In practice, when the expressions of $u_i(\bar t)$, ($i=0,1)$ are given,
we can substitute them into the above Eq. (\ref{f5}),
then intuitively we can check whether it has a solution.
Note that the expressions of $u_i(\bar{t})$, $(i=0, 1)$ determine
whether the system is solvable.
If it is solvable, then we can obtain the solution
$\bar{\Phi}_{ij}(s_0, {\bf u}, t$),  $(i,j=0, 1)$.
Note that the results in \cite{Gu2} can be regarded as a special case that has a unique solution.

\subsection{Recursive Formulas for Extracting Hidden Process}

For $\widetilde{\omega}_t \in \mathcal{F}_t$,
we can express $\widetilde{\omega}_t$ in a more clear way as follows:
$$
\widetilde{\omega}_t=(N^Y_t, N^D_t, S_{N^Y_t}, I_{N^D_t}, T_{N^D_t})
$$
where
\begin{itemize}
\item $S_{N^Y_t}=(s_1, s_2, \ldots, s_{N^Y_t})$,
\item $I_{N^D_t}=(j_1, j_2, \ldots, j_{N^D_t})$,
\item $T_{N^D_t}=(t_1, t_2, \ldots, t_{N^D_t})$,
\item $N_t^Y$ counts the number of jumps in chain $Y$ by time $t$,
\item $N_t^D$ counts the number of defaults by time $t$,
\item$(s_1, s_2, \ldots, s_{N^Y_t})$ is the collection of ordered jump times of the chain $Y$ by time $t$,
i.e., $0< s_1 < \ldots < s_{N^Y_t} \leq t$,
\item $(t_1, t_2, \ldots, t_{N^D_t})$ is the collection of ordered default times by time $t$, i.e., $0< t_1 < \ldots < t_{N^Y_t} \leq t$,
\item $(j_1, j_2, \ldots, j_{N^D_t})$ is the collection of ordered corresponding name of defaulters by time $t$, $i.e.$, name $j_i$ defaults at time $t_i$.
\end{itemize}
Here ${\widetilde \omega}_t$ can be interpreted as the state of the stochastic dynamical system at time $t$. Given the information up to time $t$, i.e., $\mathcal{F}_t$,
we divide the time period $[0, t]$ into ($N^Y_t+N^D_t$)
sub-periods, $[0, h_1]$, $(h_1, h_2]$, \ldots, $(h_{N^Y_t+N^D_t-1}, h_{N^Y_t+N^D_t}]$.
In each of them, exactly one default or one jump in $Y$ is observed.
When there is no default or jump occurred by time $t$,
the calculation of $P(X_t=x_i \mid \mathcal{F}_t)$ can be simplified and
we shall introduce it later.

Define $\bar{I}_{N^D_t}=(1,2, \ldots, K)\backslash I_{N^D_t}$.
Suppose that $s$ and $s+\Delta s$ are two endpoints of one sub-period.
The following characterizes the computational method for
$P(X_t=x_i \mid \mathcal{F}_t)$. For $\widetilde{\omega} \in \{t_k=s+\bar{t}_k \in (s, s+\Delta s]\}$,
\begin{equation}\label{f1}{
\begin{array}{lll}
P(X_s=x_i \mid \mathcal{F}_{s+\Delta s})
&=&P(X_s=x_i \mid \mathcal{F}_s, t_k=s+\bar{t}_k , j_k=\beta)\\
&=&\displaystyle \frac{P(X_s=x_i \mid \mathcal{F}_s)\cdot \left(\sum_{l=0,1}f_{t_k}^{i,l}(s+\bar{t}_k; \beta, s, \Delta s)\right)}
{ \sum_{j=0,1}P(X_s=x_j\mid \mathcal{F}_s)\cdot \left(\sum_{l=0,1}f_{t_k}^{j,l}(s+\bar{t}_k; \beta, s, \Delta s)\right)}
\end{array}}
\end{equation}
and
\begin{equation}\label{f2}{
\begin{array}{lll}
 P(X_{s+\Delta s}=x_i \mid \mathcal{F}_{s+\Delta s})
&=&\displaystyle \sum_{j=0,1} P(X_s=x_j \mid \mathcal{F}_{s+\Delta s})P(X_{s+\Delta s}=x_i \mid \mathcal{F}_{s+\Delta s}, X_s=x_j)\\
&=&\displaystyle \sum_{j=0,1} P(X_s=x_j \mid \mathcal{F}_{s+\Delta s})\displaystyle \frac{f_{t_k}^{j,i}(s+\bar{t}_k; \beta, s, \Delta s)}{\sum_{l=0,1}f_{t_k}^{j,l}(s+\bar{t}_k; \beta, s, \Delta s)}
\end{array}}
\end{equation}
where
$$
f_{t_k}^{j,i}(t; \beta, s, \Delta s){dt}
= { P(t_k \in dt, j_k=\beta, X_{s+\Delta s}=x_i \mid X_s=x_j, N_s^D, N_s^Y, I_{N_s^D})}.
$$
Similarly, we have for $\widetilde{\omega} \in \{s_k=s+\bar{s}_k \in (s, s+\Delta s]\}$,
\begin{equation}\label{f3}
\begin{array}{lll}
P(X_s=x_i \mid \mathcal{F}_{s+\Delta s})
&=&\displaystyle \frac{P(X_s=x_i \mid \mathcal{F}_s)\left(\sum_{l=0,1}f_{s_k}^{i,l}(s+\bar{s}_k; s, \Delta s\right)}{\displaystyle \sum_{j=0,1}P(X_s=x_j\mid \mathcal{F}_s)\left(\sum_{l=0,1}f_{s_k}^{j,l}(s+\bar{s}_k; s, \Delta s\right)}
\end{array}
\end{equation}
and
\begin{equation}\label{f4}
\begin{array}{lll}
P(X_{s+\Delta s}=x_i \mid \mathcal{F}_{s+\Delta s})
= &\displaystyle \sum_{j=0,1} P(X_s=x_j \mid \mathcal{F}_{s+\Delta s})\displaystyle \frac{f_{s_k}^{j,i}(s+\bar{s}_k; s, \Delta s)}{\left(\sum_{l=0,1}f_{s_k}^{j,l}(s+\bar{s}_k; s, \Delta s\right)}
\end{array}
\end{equation}
where
$$
\begin{array}{lll}
f_{s_k}^{j,i}(t; s, \Delta s) {dt}
&=&\displaystyle {P(s_k \in dt, X_{s+\Delta s}=x_i \mid X_s=x_j, N_s^D, N_s^Y, I_{N_s^D})}.
\end{array}
$$
Combining Eqs. (\ref{f1}), (\ref{f2}), (\ref{f3}) and (\ref{f4}),
we obtain a recursive method for computing
$P(X_t=x_i \mid \mathcal{F}_t)$ in terms of $f_{t_k}^{j,i}(s+\bar{t}_k; \beta, s, \Delta s)$
and $f_{s_k}^{j,i}(s+\bar{s}_k; s, \Delta s)$.
That is to say, with the fact that $P(X_0=x_0|\mathcal{F}_0)=1$ and $P(X_0=x_1|\mathcal{F}_0)=0$, we can apply them to Eq. (\ref{f1}) or Eq. (\ref{f3})
according to $\mathcal{F}_t$, and then to get $P(X_0=x_i|\mathcal{F}_{\Delta s})$ which are unknown in
the calculation of $P(X_{\Delta s}=x_i|\mathcal{F}_{\Delta s})$ in
Eq. (\ref{f2}) or Eq. (\ref{f4}).
The equation to calculate $P(X_{\Delta s}=x_i|\mathcal{F}_{\Delta s})$ should be chosen according to $\mathcal{F}_t$ as well.
By repeating this recursion procedure, we can obtain the desired conditional probabilities.


To get the expressions for the desired $f_{t_k}^{j,i}(s+\bar{t}_k; \beta, s, \Delta s)$ and $f_{s_k}^{j,i}(s+\bar{s}_k; s, \Delta s)$, we need to use the method introduced in section \ref{Preliminary}.
Replace ${\bf u}$ by $-(\eta_i(x_0), \eta_i(x_1))$, $i=0,1$ and we know that 
there exists unique solutions for $\Psi_{ij}, i,j=0,1$. Replace ${\bf u}(\bar{t})$ by $-(\lambda_i(x_0), \lambda_i(x_1))$, $i=1, \cdots, K$ in Eq. (\ref{f5}), we then could have a direct sense of whether it is solvable or not. If it is solvable and has an analytical solution, then from the definition of $f_{t_k}^{j,i}(s+\bar{t}_k; \beta, s, \Delta s)$ and $f_{s_k}^{j,i}(s+\bar{s}_k; s, \Delta s)$, we get
\begin{displaymath}{\small
\begin{array}{lll}
f_{s_k}^{j,i}(s+\bar{s}_k; s, \Delta s)&=
& \displaystyle \sum_{l=0,1} P_{jl}(\bar{s}_k) P_{li}(\Delta s-\bar{s}_k)\eta_{C(N^Y_s)}(x_l)\\
&& \times \Psi_{j l}\left(-(\eta_{C(N^Y_s)}(x_0), \eta_{C(N^Y_s)}(x_1))^T, \bar{s}_k\right)\\
&& \times \Psi_{l i}\left(-(\eta_{C(N^Y_s+1)}(x_0), \eta_{C(N^Y_s+1)}(x_1))^T, \Delta s-\bar{s}_k\right)\\
&& \times \displaystyle \bar{\Psi}_{j l}\left(s, -\sum_{i\in \bar{I}_{N^D_s}}(\lambda_i(\bar{t}|I_{N^D_s}, T_{N^D_s}, x_0), \lambda_i(\bar{t}|I_{N^D_s}, T_{N^D_s}, x_1))^T, \bar{s}_k\right)\\
&& \times \displaystyle \bar{\Psi}_{l i}\left(s+\bar{s}_k, -\sum_{i\in \bar{I}_{N^D_s}}(\lambda_i(\bar{t}|I_{N^D_s}, T_{N^D_s}, x_0), \lambda_i(\bar{t}|I_{N^D_s}, T_{N^D_s}, x_1))^T, \Delta s-\bar{s}_k\right),
\end{array}}
\end{displaymath}
\begin{displaymath}{\small
\begin{array}{lll}
f_{t_k}^{j,i}(s+\bar{t}_k; \beta, s, \Delta s)&=
& \displaystyle \sum_{l=0,1} P_{jl}(\bar{t}_k) P_{li}(\Delta s-\bar{t}_k)\lambda_\beta(s+\bar{t_k}|I_{N^D_s}, T_{N^D_s}, X_s=x_l)\\
&& \times \Psi_{j l}(-(\eta_{C(N^Y_s)}(x_0), \eta_{C(N^Y_s)}(x_1))^T, \bar{t}_k)\\
&& \times \Psi_{l i}\left(-(\eta_{C(N^Y_s)}(x_0), \eta_{C(N^Y_s)}(x_1))^T, \Delta s-\bar{t}_k\right)\\
&& \times \displaystyle \bar{\Psi}_{j l}\left(s, -\sum_{i\in \bar{I}_{N^D_s}}(\lambda_i(\bar{t}|I_{N^D_s}, T_{N^D_s}, x_0), \lambda_i(\bar{t}|I_{N^D_s}, T_{N^D_s}, x_1))^T, \bar{t}_k\right)\\
&&\times {\bar{\Psi}_{li}\left(s+\bar{t}_k, -\sum_{i\in \bar{I}_{N^D_s}^*}(\lambda_i(\bar{t}|I_{N^D_s}^*, T_{N^D_s}^*, x_0), \lambda_i(\bar{t}|I_{N^D_s}^*, T_{N^D_s}^*, x_1))^T, \Delta s-\bar{t}_k\right)}
\end{array}}
\end{displaymath}
where $I_{N^D_s}^*=I_{N^D_s}\bigcup\{\beta\}, T_{N^D_s}^*=T_{N^D_s}\bigcup\{t_{\beta}\}$ and
$$
C(x)=\left\{
\begin{array}{ll}
1, & x+Y_0 \equiv 0 \ ({\rm mod}\ 2)\\
0, & x+Y_0 \equiv 1 \ ({\rm mod}\ 2).
\end{array}
\right.
$$
If up to time $t$, no jump or default has been observed, then we have the following:
for $\widetilde{\omega} \in$ $\{$no jump or default observed in $[0,t]\}$,
$$
\begin{array}{lll}
 \displaystyle P(X_t=x_i \mid \mathcal{F}_t)
&=&\displaystyle \frac{P(X_t=x_i, {\rm no \ jump \ or \ default \ in} [0,t])}{\displaystyle \sum_{j=0,1}P(X_t=x_j, {\rm no \ jump \ or \ default \ in} [0,t])}
\end{array}
$$
where
$$
\begin{array}{lll}
P(X_t=x_j, {\rm no \ jump \ or \ default \ in} [0,t])
&= &P(X_t=x_j)\Psi_{0j}\left(-(\eta_{C(0)}(x_0), \eta_{C(0)}(x_1))^T, t\right)
 \times \bar{\Psi}_{0j}(0, \\
 &&-\sum_{i\in I}(\lambda_i(\bar{t}|I_{N^D_0}, T_{N^D_0}, x_0), \lambda_i(\bar{t}|I_{N^D_0}, T_{N^D_0}, x_1))^T, t).
\end{array}
$$
Note that if the jump intensities of chain $Y$: $\eta_i$ ($i=0, 1$), are not as simple as in our assumptions and they are also related with time directly, i.e., $\eta_i (\bar t)$,
all the algorithms introduced above are still applicable and we just need to replace $\Psi_{ij}\left(-(\eta_{C(0)}(x_0), \eta_{C(0)}(x_1))^T, t\right)$ by $\bar{\Psi}_{ij}\left(-(\eta_{C(0)}(x_0), \eta_{C(0)}(x_1))^T, t\right)$, $i,j=0,1$.
This replacement holds only when Eq. (\ref{f5})
given ${\bf u}(\bar{t})=-(\eta_0 (\bar t),\eta_1 (\bar t))$ has an analytical solution.

If Eq. (\ref{f5}) does not admit an analytical solution given $u_i(\bar{t})$, ($i=0, 1$),
we also provide numerical method in Section \ref{Numerical Method}.
Now we know how to get $P(X_t=x_i|\mathcal{F}_t)$.

\section{Default Distributions}\label{Default Distributions}

We derive the default distributions in this section.
Besides deriving closed-form expressions for default distributions, extended total hazard construction method for hidden Markov model to derive the joint default distribution is also presented.

\subsection{Closed-Form Expressions for Default Distributions}

In this subsection, we compute the conditional joint distribution of default times
$$
P(\tau^1>t^1, \tau^2>t^2, \ldots, \tau^K>t^K \mid \mathcal{F}_t)
$$
and the distribution of ordered default times
$$
P(\tau^k>s \mid \mathcal{F}_t), \quad k=1,2, \ldots, K.
$$
Notice that when $t=0$, we don't have any information, the above
two conditional probabilities become unconditional probabilities.
As for the first probability, due to the Markov property of $X_t$ and the structure of $\lambda_i(t)$, we have
$$
\begin{array}{ll}
&P(\tau^1 > t^1, \tau^2 >t^2, \ldots, \tau^K >t^K \mid \mathcal{F}_t)\\
=&\displaystyle \sum_{i=0,1}P(\tau^1 >t^1, \tau^2 >t^2, \ldots, \tau^K >t^K \mid \mathcal{F}_t^N, X_t=x_i)\times P(X_t=x_i \mid \mathcal{F}_t).
\end{array}
$$
Since we know how to calculate  $P(X_t=x_i \mid \mathcal{F}_t)$,
we only need to compute the conditional joint probability
$P(\tau^1 >t^1, \tau^2 >t^2, \ldots, \tau^K >t^K \mid \mathcal{F}_t^N, X_t=x_i)$.\\

Assume we first enter the market immediately after the $N_t^D$th default of the $K$ obligors
at time $t$, to simplify the notations, we denote $m=N_t^D$, that means we already know the information $T_m=(t_1, \cdots, t_m), I_m=(j_1, \cdots, j_m)$ and $\mathcal{F}_t^Y$ by time $t$. Then we can get the following equation:
$$
\begin{array}{ll}
&P(\tau^1 >t^1, \tau^2 >t^2, \ldots, \tau^K >t^K \mid \mathcal{F}_t^N, X_t=x_i)\\
=&P(\tau^{j_{m+1}} >t^{j_{m+1}}, \ldots, \tau^{j_K}>t^{j_K} \mid \tau^{j_1}=t_1, \ldots, \tau^{j_{m}}=t_m, X_t=x_i).
\end{array}
$$
Furthermore, we also know the relationship that
$$
f(t^{j_{m+1}}, \ldots, t^{j_{K}}\mid \mathcal{F}_t)=(-1)^{K-m}\frac{d^{K-m}}{dt^{j_{m+1}}\ldots dt^{K}}P(\tau^1>t^1, \tau^2>t^2, \ldots, \tau^K>t^K \mid \mathcal{F}_t^N, X_t=x_i)
$$
where $f(t^{j_{m+1}}, \ldots, t^{j_{K}}\mid \mathcal{F}_t)$ is the conditional joint density function. Therefore, to obtain the desired conditional probability, it suffices to find its conditional joint density function.\\

Here we employ the approach introduced by Yu (2007) \cite{Yu} (called the total hazard construction method) to derive the conditional density function.
\begin{proposition}
The expression of the density function that we intend to get is in the form of expectation
$$
f(t^{j_{m+1}}, \ldots, t^{j_{K}}\mid \mathcal{F}_t)=E\left[\sum_{l=m+1}^K\sum_{i\in\bar{I}_l}\lambda_i(t^{j_l}|I_l, T_l, X_{t^{j_l}})\cdot \exp\left(-\sum_{l=m+1}^K(\sum_{i\in\bar{I}_l}\int_{t_l}^{t^{j_l}}\lambda_i(u|I_l, T_l, X_u)du)\right)\right].
$$
\end{proposition}

\begin{proof}
Without loss of generality, we assume that $t^{j_{m+1}} < \ldots < t^{j_K}$. In this case, $\tau^{m+1}-\tau^m$ would be the first default time we observed after entering the market. By using the total hazard construction method pioneered by Yu (2007) \cite{Yu} with the information already known, we draw a collection of independent standard exponential random variables: $(E_{j_{m+1}}, \cdots, E_{j_K})$. Then we know
\begin{eqnarray*}
\tau^{m+1}-\tau^m=\min_{i\in\bar{I}_m}\Lambda_i^{-1}(E_i)=\min_{i\in\bar{I}_m}\inf\{s\ge0: \Lambda_i(s)\ge E_i\}
\end{eqnarray*}
which implies that
$$
P\left(\tau^{m+1}-\tau^m>t\mid \mathcal{F}_{\tau^m}\right)
=P\left(\min_{i\in\bar{I}_m}\inf\{s\ge0: \Lambda_i(s)\ge E_i\}>t\right).
$$
Suppose the information $\mathcal{F}_{\infty}^X$ is known, then
\begin{eqnarray*}
P(\tau^{m+1}-\tau^m>t\mid \mathcal{F}_{\tau^m})&=&\prod_{i\in\bar{I}_m}P\left(E_i>\int_{t_m}^{t_m+t}\lambda_i(u|I_m, T_m, X_u)du\right)\\
&=&\prod_{i\in\bar{I}_m}\exp\left(-\int_{t_m}^{t_m+t}\lambda_i(u|I_m, T_m, X_u)du\right)\\
&=&\exp\left(-\sum_{i\in\bar{I}_m}\int_{t_m}^{t_m+t}\lambda_i(u|I_m, T_m, X_u)du\right).
\end{eqnarray*}

Then if we assume that $\tau^m<t<\tau^{m+1}$ and $t^i>\tau^i, i=1, \ldots, m$, and let $\lambda^{m+1}(t)$ denote the $(m+1)$th default rate at time $t$, then
$$
\lambda^{m+1}(t)=\sum_{i\in\bar{I}_m}\int_{t_m}^{t}\lambda_i(u|I_m, T_m, X_u)du.
$$
Since
$$
P(\tau^{m+1}>t\mid \mathcal{F}_{\tau^m}, X_{s(t_m<s<\infty)})=e^{-\sum_{i\in\bar{I}_m}\int_{t_m}^{t}\lambda_i(u|I_m, T_m, X_u)du}=e^{-\lambda^{m+1}(t)}
$$
we have
\begin{eqnarray*}
&& P(\tau^{j_{m+1}} >t^{j_{m+1}}, \ldots, \tau^{j_K}>t^{j_K}\mid \mathcal{F}_t, X_{s(t_m<s<\infty)})\\
&&=\prod_{l=m+1}^KP(\tau^{j_l} >t^{j_l}\mid \mathcal{F}_t, X_{s(t_m<s<\infty)})\\
&&=\prod_{l=m+1}^K e^{-\lambda^{l}(t^{j_l})}=\prod_{l=m+1}^K \exp\left(-\sum_{i\in\bar{I}_{l-1}}\int_{t_{l-1}}^{t^{j_l}}\lambda_i(u|I_{l-1}, T_{l-1}, X_u)du\right)\\
&&=\exp\left(-\sum_{l=m+1}^K(\sum_{i\in\bar{I}_{l-1}}\int_{t_{l-1}}^{t^{j_l}}\lambda_i(u|I_{l-1}, T_{l-1}, X_u)du)\right)
\end{eqnarray*}
and therefore
\begin{eqnarray*}
&&f(t^{j_{m+1}}, \ldots, t^{j_{K}}\mid \mathcal{F}_t, X_{s(t_m<s<\infty)})\\
&&=(-1)^{K-m}\frac{d^{K-m}}{dt^{j_{m+1}}\ldots dt^{K}}P(\tau^{j_{m+1}} >t^{j_{m+1}}, \ldots, \tau^{j_K}>t^{j_K}\mid \mathcal{F}_t, X_{s(t_m<s<\infty)})\\
&&=(-1)^{K-m}\frac{d^{K-m}}{d t^{j_{m+1}}\ldots dt^{K}}  \exp\left(-\sum_{l=m+1}^K(\sum_{i\in\bar{I}_{l-1}}\int_{t_{l-1}}^{t^{j_l}}\lambda_i(u|I_{l-1}, T_{l-1}, X_u)du)\right) \arrowvert_{t_{l-1}=t^{j_{l-1}}}\\
&&=\prod_{l=m+1}^K\sum_{i\in\bar{I}_{l-1}}\lambda_i(t^{j_l}|I_{l-1}, T_{l-1}, X_{t^{j_l}})\cdot \exp\left(-\sum_{l=m+1}^K(\sum_{i\in\bar{I}_{l-1}}\int_{t^{j_{l-1}}}^{t^{j_l}}\lambda_i(u|I_{l-1}, T_{l-1}, X_u)du)\right)
\end{eqnarray*}
and
\begin{eqnarray*}
&&f(t^{j_{m+1}}, \ldots, t^{j_{K}}\mid \mathcal{F}_t)=E[f(t^{j_{m+1}}, \ldots, t^{j_{K}} \mid \mathcal{F}_t, X_{s(t_m<s<\infty)})]\\
&=&E\left[\prod_{l=m+1}^K\sum_{i\in\bar{I}_{l-1}}\lambda_i(t^{j_l}|I_{l-1}, T_{l-1}, X_{t^{j_l}})\cdot \exp\left(-\sum_{l=m+1}^K(\sum_{i\in\bar{I}_{l-1}}\int_{t^{j_{l-1}}}^{t^{j_l}}\lambda_i(u|I_{l-1}, T_{l-1}, X_u)du)\right)\right]\\
&=&E\left[\prod_{l=m+1}^K\sum_{i\in\bar{I}_{l-1}}\lambda_i(t^{j_l}|I_{l-1}, T_{l-1}, X_{t^{j_l}})\cdot \exp\left(-\sum_{l=m+1}^K(\int_{t^{j_{l-1}}}^{t^{j_l}}\sum_{i\in\bar{I}_{l-1}}\lambda_i(u|I_{l-1}, T_{l-1}, X_u)du)\right)\right].
\end{eqnarray*}
\end{proof}

If Eq. (\ref{f5}) in the previous section given ${\bf u}(\bar{t})=-(\lambda_i(x_0), \lambda_i(x_1))$, $i=1, \ldots, K$, has unique solutions, then we further have the following result.

\begin{proposition}
The explicit formula for calculating the desired density function is in this form:
\begin{eqnarray*}
&&f(t^{j_{m+1}}, \ldots, t^{j_{K}}\mid \mathcal{F}_t)=(-1)^{K-m}\cdot\sum_{l_{m+1}=0,1}\sum_{l_{m+2}=0,1}\cdots\sum_{l_K=0,1}\\
&\cdot&\frac{d(\bar{\Psi}_{il_{m+1}}(t^{j_m},-\sum_{i\in\bar{I}_m}(\lambda_i(\bar{t}|I_m,T_m, x_0),\lambda_i(\bar{t}|I_m,T_m, x_1))^T, t^{j_{m+1}}-t^{j_m}))}{d t^{j_{m+1}}}\\
&\cdot&\frac{d(\bar{\Psi}_{l_{m+1}l_{m+2}}(t^{j_{m+1}},-\sum_{i\in\bar{I}_{m+1}}(\lambda_i(\bar{t}|I_{m+1},T_{m+1}, x_0),\lambda_i(\bar{t}|I_{m+1},T_{m+1}, x_1))^T, t^{j_{m+2}}-t^{j_{m+1}}))}{d t^{j_{m+2}}}\\
&\cdot&\cdots\cdot\frac{d(\bar{\Psi}_{l_{K-1}l_K}(t^{j_{K-1}},-\sum_{i\in\bar{I}_{K-1}}(\lambda_i(\bar{t}|I_{K-1},T_{K-1}, x_0),\lambda_i(\bar{t}|I_{K-1},T_{K-1}, x_1))^T, t^{j_K}-t^{j_{K-1}}))}{d t^{j_K}}
\end{eqnarray*}
where $\bar{\Psi}_{ij}, i, j=0, 1$ are the moment generating function defined in Section \ref{Extraction of Hidden Process with Observable Processes}.
\end{proposition}

\begin{proof}
We note that
{\small
\begin{eqnarray*}
&&E\left[\prod_{l=m+1}^K\sum_{i\in\bar{I}_{l-1}}\lambda_i(t^{j_l}|I_{l-1}, T_{l-1}, X_{t^{j_l}})\cdot e^{-\sum_{l=m+1}^K(\int_{t^{j_{l-1}}}^{t^{j_l}}\sum_{i\in\bar{I}_{l-1}}\lambda_i(u|I_{l-1}, T_{l-1}, X_u)du)}\right]\\
&=&\sum_{l_{m+1}=0,1}\sum_{l_{m+2}=0,1}\cdots\sum_{l_K=0,1}\\
&&E\left[\sum_{i\in\bar{I}_m}\lambda_i(t^{j_{m+1}}|I_m, T_m, X_{t^{j_{m+1}}})\cdot e^{\int_{t^{j_m}}^{t^{j_{m+1}}}\sum_{i\in\bar{I}_m}\lambda_i(u|I_m, T_m, X_u)du}| X_{t^{j_m}}=i, X_{t^{j_{m+1}}}=l_{m+1}\right]\\
&\cdot&E\left[\sum_{i\in\bar{I}_{m+1}}\lambda_i(t^{j_{m+2}}|I_{m+1}, T_{m+1}, X_{t^{j_{m+2}}})\cdot e^{\int_{t^{j_{m+1}}}^{t^{j_{m+2}}}\sum_{i\in\bar{I}_{m+1}}\lambda_i(u|I_{m+1}, T_{m+1}, X_u)du}| X_{t^{j_{m+1}}}=l_{m+1}, X_{t^{j_{m+2}}}=l_{m+2}\right]\\
&\cdot&\cdots\cdot E\left[\sum_{i\in\bar{I}_{K-1}}\lambda_i(t^{j_K}|I_{K-1}, T_{K-1}, X_{t^{j_K}})\cdot e^{\int_{t^{j_{K-1}}}^{t^{j_K}}\sum_{i\in\bar{I}_{K-1}}\lambda_i(u|I_{K-1}, T_{K-1}, X_u)du}| X_{t^{j_{K-1}}}=l_{K-1}, X_{t^{j_K}}=l_K\right]\\
&=&(-1)^{K-m}\cdot\sum_{l_{m+1}=0,1}\sum_{l_{m+2}=0,1}\cdots\sum_{l_K=0,1}\frac{d\left(E\left[e^{\int_{t^{j_m}}^{t^{j_{m+1}}}\sum_{i\in\bar{I}_m}\lambda_i(u|I_m, T_m, X_u)du}| X_{t^{j_m}}=i, X_{t^{j_{m+1}}}=l_{m+1}\right]\right)}{d t^{j_{m+1}}}\\
&\cdot&\frac{d \left(E\left[e^{\int_{t^{j_{m+1}}}^{t^{j_{m+2}}}\sum_{i\in\bar{I}_{m+1}}\lambda_i(u|I_{m+1}, T_{m+1}, X_u)du}| X_{t^{j_{m+1}}}=l_{m+1}, X_{t^{j_{m+2}}}=l_{m+2}\right]\right)}{d t^{j_{{m+2}}}}\\
&\cdot&\cdots\cdot\frac{d\left(E\left[e^{\int_{t^{j_{K-1}}}^{t^{j_K}}\sum_{i\in\bar{I}_{K-1}}\lambda_i(u|I_{K-1}, T_{K-1}, X_u)du}| X_{t^{j_{K-1}}}=l_{K-1}, X_{t^{j_K}}=l_K\right]\right)}{d t^{j_K}}\\
&=&(-1)^{K-m}\cdot\sum_{l_{m+1}=0,1}\sum_{l_{m+2}=0,1}\cdots\sum_{l_K=0,1}\frac{d(\bar{\Psi}_{il_{m+1}}(t^{j_m},-\sum_{i\in\bar{I}_m}(\lambda_i(\bar{t}|I_m,T_m, x_0),\lambda_i(\bar{t}|I_m,T_m, x_1))^T, t^{j_{m+1}}-t^{j_m}))}{d t^{j_{m+1}}}\\
&\cdot&\frac{d(\bar{\Psi}_{l_{m+1}l_{m+2}}(t^{j_{m+1}},-\sum_{i\in\bar{I}_{m+1}}(\lambda_i(\bar{t}|I_{m+1},T_{m+1}, x_0),\lambda_i(\bar{t}|I_{m+1},T_{m+1}, x_1))^T, t^{j_{m+2}}-t^{j_{m+1}}))}{d t^{j_{m+2}}}\\
&\cdot&\cdots\cdot\frac{d(\bar{\Psi}_{l_{K-1}l_K}(t^{j_{K-1}},-\sum_{i\in\bar{I}_{K-1}}(\lambda_i(\bar{t}|I_{K-1},T_{K-1}, x_0),\lambda_i(\bar{t}|I_{K-1},T_{K-1}, x_1))^T, t^{j_K}-t^{j_{K-1}}))}{d t^{j_K}}
\end{eqnarray*}
}
\end{proof}

Similarly if the equations related to $\bar{\Psi}_{ij}$
do not have analytical solutions, then we can use the same approximation method which will be discussed in the next section to approximate $\bar{\Psi}_{ij}$ with $\Psi_{ij}$.
Thus one can obtain an explicit approximation expression for the density function
$f(t^{j_{m+1}}, \ldots, t^{j_{K}}\mid \mathcal{F}_t)$.
When the expressions of the default intensities are homogeneous and symmetric,
$$
\begin{array}{lll}
&&P(\tau^{j_{m+1}}<\cdots<\tau^{j_k}<s<\tau^{j_{k+1}}<\cdots<\tau^{j_K}\mid \mathcal{F}_t)\\
&=&\displaystyle \int_{t_m}^{t}\int_{t^{j_{m+1}}}^{t}\cdots\int_{t^{j_{k-1}}}^{t}\int_{t}^{\infty}\int_{t^{j_{k+1}}}^{\infty}\cdots\int_{t^{j_{K-1}}}^{\infty}f(t^{j_{m+1}}, t^{j_{m+2}}, \ldots, t^{j_K}\mid \mathcal{F}_s) dt^{j_K}\cdots dt^{t_{j_{m+1}}}.
\end{array}
$$
Because they are homogeneous and symmetric,
$$
P(\tau^{j_k}\le s<\tau^{j_{k+1}}\mid \mathcal{F}_t)=(K-m)! P(\tau^{j_{m+1}}<\cdots<\tau^{j_k}<s<\tau^{j_{k+1}}<\cdots<\tau^{j_K}\mid \mathcal{F}_t).
$$
Furthermore, we have
$$
P(\tau^{j_k}>s\mid \mathcal{F}_t)=\sum_{i=m}^{k-1}P(\tau^{j_i}\le s<\tau^{j_{i+1}}\mid \mathcal{F}_t).
$$

\subsection{Extended Total Hazard Construction Method for HMM}

We further extend the total hazard construction method to make it applicable to various forms of default intensities modulated by a hidden Markov process, then to gain the joint default distribution.

The total hazard accumulated by obligor $i$ by time $t$, denoted by
$\psi_i(t|I_{N_t^D}, T_{N_t^D}, X_t)$, can be defined as follows:
\begin{equation}
\psi_i(t|I_{N_t^D}, T_{N_t^D}, X_t)=\sum_{l=0}^{N_t^D-1}\Lambda_i(t_{l+1}-t_l|I_l, T_l, X_{t_{l+1}})+\Lambda_i(t-t_{N_t^D}|I_{N_t^D}, T_{N_t^D}, X_t)
\end{equation}
where
\begin{equation}
\Lambda_i(s|I_l, T_l, X_{t_l+s})=\int_{t_l}^{t_l+s}\lambda_i(\mu|I_l, T_l, X_{\mu})d\mu
\end{equation}
is the total hazard accumulated by obligor $i$ in the time interval $[t_l, t_l+s]$. Note that the default processes
are independent unit exponential random variables.
And we define the inverse function
\begin{equation}
\Lambda_i^{-1}(x|I_k, T_k, N_{\infty}^Y, S_{N_{\infty}^Y})=\inf\{s: \Lambda_i(s|I_k, T_k, X_{t_k+s})\ge x\}, x\ge0
\end{equation}
where $(N_{\infty}^Y, S_{N_{\infty}^Y})\in \mathcal{F}_{\infty}^{Y}$
is the entire history of the path of $Y$, $N_{\infty}^Y$ is the entire number of jump in the chain $Y$ and $S_{N_{\infty}^Y}$ is the collection of corresponding ordered jump times.

The total hazard can be constructed by the following recursive procedure:\\
$$
\begin{tabular}{l}
\hline
{\bf Step 1.} Generate a complete sample path of $Y$,
and denote it as $(N_{\infty}^Y, S_{N_{\infty}^Y})\in \mathcal{F}_{\infty}^Y$.\\
Generate a collection of i.i.d. unit exponential random variables $(E_1, \cdots, E_K)$.\\

{\bf Step 2.} Let
$j_1=\arg\min\{\Lambda_i^{-1}(E_i): i=1, \cdots, K\}$ and define
$\hat{\tau}^{j_1}=\Lambda_{j_1}^{-1}(E_{j_1})$.\\
Note that $T_1=(t_1), t_1=\hat{\tau}^{j_1}, I_1=(j_1)$.\\

{\bf Step 3.} (i)
Assume that $(\hat{\tau}^{j_1}, \ldots, \hat{\tau}^{j_{m-1}})$
and the simulated path of $X_s(0\le s< \hat{\tau}^{j_{m-1}})$ are \\
already obtained as $T_{m-1}=(t_1, \ldots, t_{m-1}), t_l=\hat{\tau}^{j_l}, l=1, \ldots, m-1$
and $I_{m-1}=(j_1, \ldots, j_{m-1})$, \\
where $m\ge 2$. By using the conditional probability of \\
$P(X_s=x_i|\tilde{\mathcal{F}}_s), \quad i=0, 1, x_0=0, x_1=1, s\ge\hat{\tau}^{j_{m-1}}$
and $\tilde{\mathcal{F}}_s=\mathcal{F}_s^Y\vee T_{m-1}\vee I_{m-1}$,\\
we can generate a sequence of random numbers of $X_s$ under this conditional probability.\\
We can then obtain the simulated path of $X_s, s\ge\hat{\tau}^{j_{m-1}}$ which will be useful
in the calculation of \\
$\Lambda_i^{-1}(x|I_{m-1}, T_{m-1}, N_{\infty}^Y, S_{N_{\infty}^Y})$. \\

(ii) Note that $\bar{I}_{m-1}=(1,2, \ldots, K)\backslash I_{m-1}$.\\
Therefore, with the information of $T_{m-1}, I_{m-1}$ and the path of $X_s(0\le s< \hat{\tau}^{j_{m-1}})\cup X_s(s\ge\hat{\tau}^{j_{m-1}})$, \\
i.e., the path of $X$. We let\\
$j_m=\arg\min\{\Lambda_i^{-1}(E_i-\psi_i(t_{m-1}|I_{m-1}, T_{m-1}, X_{t_{m-1}})|I_{m-1}, T_{m-1}, N_{\infty}^Y, S_{N_{\infty}^Y}): i\in \bar{I}_{m-1}\}$\\
where $\psi_i(t_{m-1}|I_{m-1}, T_{m-1}, X_{t_{m-1}})$ is the total hazard accumulated by Name $i$ under the condition \\
of defaults and information of chain $X$ by the $(m-1)$th default time, i.e., $t_{m-1}$.\\
Then we let\\
$\hat{\tau}^{j_m}=t_{m-1}+\Lambda_{j_m}^{-1}(E_{j_m}-\psi_{j_m}(t_{m-1}|I_{m-1}, T_{m-1}, X_{t_{m-1}})|I_{m-1}, T_{m-1}, N_{\infty}^Y, S_{N_{\infty}^Y})$\\
and reserve the simulated path of
$X_s, \hat{\tau}^{j_{m-1}}\le s< \hat{\tau}^{j_m}$ at this step.\\
Thus,  with the simulated path, we can get the simulated path of $X_s, 0\le s<\hat{\tau}^{j_m}$.\\

{\bf Step 4.} If $m=K$, then stop. Otherwise, increase $m$ by $1$ and go to {\bf Step 3}.\\
\hline\\
\end{tabular}
$$

From the recursive procedure, we can obtain the distribution of $\hat{\tau}$.
According to Shaked and Shanthikumar (1987) \cite{Shaked} and Yu (2007) \cite{Yu},
the distribution of $\hat{\tau}$ obtained from the above recursive processes
is equal to the distribution of the original default time $\tau$.
This gives the following results.

\begin{proposition}
Let $\tau$ be the default time with the intensities
$$
\lambda_t^i=\lambda_i(t|I_{N_t^D}, T_{N_t^D}, X_t), \quad i=1,\ldots, K
$$
and the related jump processes satisfying the assumptions mentioned in Section 2.
Construct $\hat{\tau}$ according to Steps  $1-4$ with the intensity equal to
$$
\lambda_i(t|I_{N_t^D}, T_{N_t^D}, X_t), \quad i=1,2,\ldots, K.
$$
Let $\mathcal{F}_t'$ be the minimal filtration containing $\mathcal{F}^Y_t$ and the information of the default processes related to $\hat{\tau}$ by time $t$, and $P'$ be the distribution of $(Y, \hat{\tau})$.
Then every element in $\hat{\tau}^i$ has $(P', \mathcal{F}_t')$-intensity of the form:
$$
\lambda_i(t|I_{N_t^D}, T_{N_t^D}, X_t), \quad i=1,2,\ldots, K.
$$
\end{proposition}
Therefore, we can generate $\tau$ by just generating $\hat{\tau}$.

\section{Numerical Approximation Method}\label{Numerical Method}

In this section, we consider an outstanding problem in Section
\ref{Extraction of Hidden Process with Observable Processes}.
If Eq. (\ref{f5}) does not admit an analytical solution given $u_i(\bar{t})$, ($i=0, 1$)
then we shall try to use another method to approximate the conditional probability $P(X_t=x_i|\mathcal{F}_t)$.
We can consider approximating $\bar{\Psi}_{ij}(s_0, {\bf u}, t)$ directly.
As we mentioned before, it is because of the default intensities
$\lambda_i$, ($i=1, \ldots, K$) which give
Eq. (\ref{f5}) with
$$
{\bf u}(\bar{t})=-(\lambda_i(x_0), \lambda_i(x_1)), \quad i=1, \ldots, K
$$
does not have an analytical solution,
and hence we cannot obtain closed-form expressions for $\bar{\Psi}_{ij}(s_0, {\bf u}, t)$.
Thus, we need to approximate the moment generating function
$\bar{\Psi}_{ij}(s_0, {\bf u}, t)$ when the default intensities are applied.
If the error of $\bar{\Psi}_{ij}(s_0, {\bf u}, t)$ is less than any arbitrary $\epsilon$
then according to the expression of $f_{s_k}^{j,i}(s+\bar{s}_k; s, \Delta s)$
and $f_{t_k}^{j,i}(s+\bar{t}_k; \beta, s, \Delta s)$ given below, we know that their relative errors can be controlled.
Furthermore, from the recursive method for $P(X_t=x_i|\mathcal{F}_t)$ presented in Section \ref{Extraction of Hidden Process with Observable Processes},
the error of this conditional probability may be controlled.

In the following, we are going to illustrate how the approximation works.
When the length of the time interval length is small enough,
without loss of generality, we can approximately assign $t$ in the default intensities
$\lambda_i(t)$ to be the left value of the concerned time interval,
i.e., $t=s_0$ when the time interval is $[s_0, s_0+\Delta\bar{s}]$.
Then we can still apply the moment generating function given ${\bf u}(\bar{t})={\bf u}(s_0)=\bar{{\bf u}}$, and we know the corresponding Eq. (\ref{f5}) has a unique solution. But we need to ensure that by using this method,
the error of $\bar{\Psi}_{ij}(s_0, {\bf u}, \Delta\bar{s})$ can be controlled such that it can be less
than any arbitrarily given $\epsilon$.

\begin{proposition}
The error control $\Delta\Psi_{ij}(s_0, {\bf u}, \Delta\bar{s})<\epsilon<1$, where $\epsilon$ is arbitrary, can be achieved by requiring $\Delta\bar{s}$ to satisfy
$$
\Delta\bar{s}<\frac{-\ln(1-\epsilon)}{K\cdot\lambda_{\max}(s_0)}
$$
where
$$
\lambda_{\max}(s_0)=\max\limits_{i=1,\ldots, K}\{\lambda_i(s), s\in [0, s_0]\}
$$
and
$$
\Delta\Psi_{ij}(s_0, {\bf u}, \Delta\bar{s})=|\Psi_{ij}(\bar{\bf u}, \Delta\bar{s})-\bar{\Psi}_{ij}(s_0, {\bf u}, \Delta\bar{s})|
$$
and
$$
\bar{\bf u}(t)={\bf u}(\tilde{s}_{k-1}) \quad {\rm for} \quad t\in(\tilde{s}_{k-1}, \tilde{s}_k]
$$
and
$$
[0, s_0]=[\tilde{s}_0, \tilde{s}_1]\bigcup(\tilde{s}_1, \tilde{s}_2]\bigcup\cdots\bigcup(\tilde{s}_{n-1}, \tilde{s}_n].
$$

\end{proposition}

\begin{proof}
Note that there are $K$ entities, so when the default intensity is applied, i.e.,
$$
{\bf u}(\bar{t})=-(\lambda_i(x_0), \lambda_i(x_1))
\quad {\rm or} \quad {\bf u}=-(\lambda_i(x_0), \lambda_i(x_1)), i=1, \ldots, K,
$$
we notice the relationships that
$$
E[e^{-K\cdot\lambda_{\max}(s_0)\cdot \Delta\bar{s}}]\le\Psi_{ij}(\bar{\bf u}, \Delta\bar{s})\le E[e^{K\cdot0\cdot \Delta\bar{s}}]
$$
and
$$
E[e^{-K\cdot\lambda_{\max}(s_0)\cdot \Delta\bar{s}}]\le\bar{\Psi}_{ij}(s_0, {\bf u}, \Delta\bar{s})\le E[e^{K\cdot0\cdot \Delta\bar{s}}].
$$
Since all $\lambda_i, i=1, \ldots, K$ are nonnegative,
therefore, we have the following relationship:
$$
\Delta\Psi_{ij}(s_0, {\bf u}, \Delta\bar{s})\le E[e^{K\cdot0\cdot \Delta\bar{s}}-e^{-K\cdot\lambda_{\max}(s_0)\cdot \Delta\bar{s}}]<\epsilon
$$
if and only if
$$
e^{K\cdot\lambda_{\max}(s_0)\cdot \Delta\bar{s}} < \frac{1}{1-\epsilon}
$$
if and only if
$$
\Delta\bar{s} < \frac{-\ln(1-\epsilon)}{K\cdot\lambda_{\max}(s_0)}.
$$
\end{proof}

We can simply let $\Delta\bar{s}=\frac{-\ln(1-\epsilon)}{K\cdot\lambda_{\max}(s_0)}$,
it is enough to make the error of $\bar{\Psi}_{ij}(s_0, {\bf u}, \Delta\bar{s})$ controllable.
Here we are in the position to approximate $f_{s_k}^{j,i}(s+\bar{s}_k; s, \Delta s)$.
First, we partition the time interval $[s, s+\bar{s}_k]$ evenly with step size equal to
$\Delta\bar{s}=\frac{-\ln(1-\epsilon)}{K\cdot\lambda_{\max}(s+\Delta s)}$, and denote
$M_1(s, \Delta\bar{s})=\left[\frac{\bar{s}_k}{\Delta\bar{s}}\right]$.
That is to say,
$$
[s, s+\bar{s}_k]=[s, s+\Delta\bar{s}]\bigcup[s+\Delta\bar{s}, s+2\Delta\bar{s}]\bigcup\cdots \bigcup[s+M_1(s, \Delta\bar{s})\cdot\Delta\bar{s}, \bar{s}_k]
$$
Moreover, we do the same thing for the remaining time interval:
$[s+\bar{s}_k, \Delta s]$ and denote $M_2(s, \Delta\bar{s})=\left[\frac{\Delta s-\bar{s}_k}{\Delta\bar{s}}\right]$.
We denote $M_1=M_1(s, \Delta\bar{s})$ and $M_2=M_2(s, \Delta\bar{s})$.
Now the explicit approximation formula is given as follows:
\begin{displaymath}{
\begin{array}{ll}
& f_{s_k}^{j,i}(s+\bar{s}_k; s, \Delta s)=\\
& \displaystyle \sum_{l=0,1}\sum_{l_1=0,1}\cdots\sum_{l_{M_1}=0,1}\sum_{\bar{l}_1=0,1}\cdots\sum_{\bar{l}_{M_2}=0,1} P_{jl}(\bar{s}_k) P_{li}(\Delta s-\bar{s}_k)\eta_{C(N^Y_s)}(x_l)\\
& \times \Psi_{j l}\left(-(\eta_{C(N^Y_s)}(x_0), \eta_{C(N^Y_s)}(x_1))^T, \bar{s}_k\right)\\
& \times \Psi_{l i}\left(-(\eta_{C(N^Y_s+1)}(x_0), \eta_{C(N^Y_s+1)}(x_1))^T, \Delta s-\bar{s}_k\right)\\
& \times \displaystyle \Psi_{j l_1}\left(-\sum_{i\in \bar{I}_{N^D_s}}(\lambda_i(s|I_{N^D_s}, T_{N^D_s}, x_0), \lambda_i(s|I_{N^D_s}, T_{N^D_s}, x_1))^T, \Delta\bar{s}\right)\\
& \times \displaystyle \Psi_{l_1 l_2}\left(-\sum_{i\in \bar{I}_{N^D_s}}(\lambda_i(s+\Delta\bar{s}|I_{N^D_s}, T_{N^D_s}, x_0), \lambda_i(s+\Delta\bar{s}|I_{N^D_s}, T_{N^D_s}, x_1))^T, \Delta\bar{s}\right)\\
& \times \cdots\\
& \times \displaystyle \Psi_{l_{M_1} l}\left(-\sum_{i\in \bar{I}_{N^D_s}}(\lambda_i(s+M_1\cdot\Delta\bar{s}|I_{N^D_s}, T_{N^D_s}, x_0), \lambda_i(s+M_1\cdot\Delta\bar{s}|I_{N^D_s}, T_{N^D_s}, x_1))^T, \bar{s}_k-s-M_1\cdot\Delta\bar{s}\right)\\
& \times \displaystyle \Psi_{l \bar{l}_1}\left(-\sum_{i\in \bar{I}_{N^D_s}}(\lambda_i(\bar{s}_k|I_{N^D_s}, T_{N^D_s}, x_0), \lambda_i(\bar{s}_k|I_{N^D_s}, T_{N^D_s}, x_1))^T, \Delta\bar{s}\right)\\
& \times \displaystyle \Psi_{\bar{l}_1 \bar{l}_2}\left(-\sum_{i\in \bar{I}_{N^D_s}}(\lambda_i(\bar{s}_k+\Delta\bar{s}|I_{N^D_s}, T_{N^D_s}, x_0), \lambda_i(\bar{s}_k+\Delta\bar{s}|I_{N^D_s}, T_{N^D_s}, x_1))^T, \Delta\bar{s}\right)\\
& \times \cdots\\
& \times \displaystyle \Psi_{\bar{l}_{M_2} i}\left(-\sum_{i\in \bar{I}_{N^D_s}}(\lambda_i(\bar{s}_k+M_2\cdot\Delta\bar{s}|I_{N^D_s}, T_{N^D_s}, x_0), \lambda_i(\bar{s}_k+M_2\cdot\Delta\bar{s}|I_{N^D_s}, T_{N^D_s}, x_1))^T, \Delta s-\bar{s}_k-M_2\cdot\Delta\bar{s}\right).
\end{array}}
\end{displaymath}
Similarly, we can get
\begin{displaymath}{\small
\begin{array}{ll}
& f_{t_k}^{j,i}(s+\bar{t}_k; \beta, s, \Delta s)=\\
& \displaystyle \sum_{l=0,1}\sum_{l_1=0,1}\cdots\sum_{l_{\bar{M}_1}=0,1}\sum_{\bar{l}_1=0,1}\cdots\sum_{\bar{l}_{\bar{M}_2}=0,1} P_{jl}(\bar{t}_k) P_{li}(\Delta s-\bar{t}_k)\lambda_\beta(s+\bar{t}_k|I_{N^D_s}, T_{N^D_s}, x_l)\\
& \times \Psi_{j l}\left(-(\eta_{C(N^Y_s)}(x_0), \eta_{C(N^Y_s)}(x_1))^T, \bar{t}_k\right)\\
& \times \Psi_{l i}\left(-(\eta_{C(N^Y_s)}(x_0), \eta_{C(N^Y_s)}(x_1))^T, \Delta s-\bar{t}_k\right)\\
& \times \displaystyle \Psi_{j l_1}\left(-\sum_{i\in \bar{I}_{N^D_s}}(\lambda_i(s|I_{N^D_s}, T_{N^D_s}, x_0), \lambda_i(s|I_{N^D_s}, T_{N^D_s}, x_1))^T, \Delta\bar{s}\right)\\
& \times \displaystyle \Psi_{l_1 l_2}\left(-\sum_{i\in \bar{I}_{N^D_s}}(\lambda_i(s+\Delta\bar{s}|I_{N^D_s}, T_{N^D_s}, x_0), \lambda_i(s+\Delta\bar{s}|I_{N^D_s}, T_{N^D_s}, x_1))^T, \Delta\bar{s}\right)\\
& \times \cdots\\
& \times \displaystyle \Psi_{l_{\bar{M}_1} l}\left(-\sum_{i\in \bar{I}_{N^D_s}}(\lambda_i(s+\bar{M}_1\cdot\Delta\bar{s}|I_{N^D_s}, T_{N^D_s}, x_0), \lambda_i(s+\bar{M}_1\cdot\Delta\bar{s}|I_{N^D_s}, T_{N^D_s}, x_1))^T, \bar{t}_k-s-\bar{M}_1\cdot\Delta\bar{s}\right)
\end{array}}
\end{displaymath}
\begin{displaymath}{\small
\begin{array}{ll}
& \times \displaystyle \Psi_{l \bar{l}_1}\left(-\sum_{i\in \bar{I}_{N^D_s}^*}(\lambda_i(\bar{t}_k|I_{N^D_s}^*, T_{N^D_s}^*, x_0), \lambda_i(\bar{t}_k|I_{N^D_s}^*, T_{N^D_s}^*, x_1))^T, \Delta\bar{s}\right)\\
& \times \displaystyle \Psi_{\bar{l}_1 \bar{l}_2}\left(-\sum_{i\in \bar{I}_{N^D_s}^*}(\lambda_i(\bar{t}_k+\Delta\bar{s}|I_{N^D_s}^*, T_{N^D_s}^*, x_0), \lambda_i(\bar{t}_k+\Delta\bar{s}|I_{N^D_s}^*, T_{N^D_s}^*, x_1))^T, \Delta\bar{s}\right)\\
& \times \cdots\\
& \times \displaystyle \Psi_{\bar{l}_{\bar{M}_2} i}\left(-\sum_{i\in \bar{I}_{N^D_s}^*}(\lambda_i(\bar{s}_k+\bar{M}_2\cdot\Delta\bar{s}|I_{N^D_s}^*, T_{N^D_s}^*, x_0), \lambda_i(\bar{s}_k+\bar{M}_2\cdot\Delta\bar{s}|I_{N^D_s}^*, T_{N^D_s}^*, x_1))^T, \Delta s-\bar{t}_k-\bar{M}_2\Delta\bar{s}\right)
\end{array}}
\end{displaymath}
where $\bar{M}_1=\left[\frac{\bar{t}_k}{\Delta\bar{s}}\right]$ and $\bar{M}_2=\left[\frac{\Delta s-\bar{t}_k}{\Delta\bar{s}}\right]$.\\
$$
\begin{array}{ll}
&P(X_t=x_j, {\rm no \ jump \ or \ default \ in} [0,t])\\
= & \displaystyle\sum_{l_1=0,1}\sum_{l_2=0,1}\cdots\sum_{l_{M}=0,1}P(X_t=x_j)\Psi_{0j}\left(-(\eta_{C(0)}(x_0), \eta_{C(0)}(x_1))^T, t\right)\\
& \times \Psi_{0 l_1}(-\sum_{i\in I}(\lambda_i(0|I_{N^D_0}, T_{N^D_0}, x_0), \lambda_i(0|I_{N^D_0}, T_{N^D_0}, x_1))^T, \Delta\bar{s})\\
& \times \Psi_{l_1 l_2}(-\sum_{i\in I}(\lambda_i(\Delta\bar{s}|I_{N^D_0}, T_{N^D_0}, x_0), \lambda_i(\Delta\bar{s}|I_{N^D_0}, T_{N^D_0}, x_1))^T, \Delta\bar{s})\\
& \times \cdots\\
& \times \Psi_{l_{M} j}(-\sum_{i\in I}(\lambda_i(M\cdot\Delta\bar{s}|I_{N^D_0}, T_{N^D_0}, x_0), \lambda_i(M\cdot\Delta\bar{s}|I_{N^D_0}, T_{N^D_0}, x_1))^T, t-M\cdot\Delta\bar{s})
\end{array}
$$
where $M=\left[\frac{t}{\Delta\bar{s}}\right]$.

Now we know how to ensure $\Delta\Psi_{ij}(s_0, {\bf u}, \Delta\bar{s})<\epsilon$, and have the formulas for calculating $f_{s_k}^{j,i}(s+\bar{s}_k; s, \Delta s)$, $f_{t_k}^{j,i}(s+\bar{t}_k; \beta, s, \Delta s)$ and $P(X_t=x_j, {\rm no \ jump \ or \ default \ in} [0,t])$.
We can then discuss how to choose $\epsilon$ such that the relative error of them can be controlled as small as we wish, i.e., $\zeta$.
Taking $f_{s_k}^{j,i}(s+\bar{s}_k; s, \Delta s)$ as an example in the following discussion, the results related to the others are similar.

\begin{proposition}
To ensure the relative error of $f_{s_k}^{j,i}(s+\bar{s}_k; s, \Delta s)$, i.e.,
$$
\frac{|\bar{f}_{s_k}^{j,i}(s+\bar{s}_k; s, \Delta s)-f_{s_k}^{j,i}(s+\bar{s}_k; s, \Delta s)|}{f_{s_k}^{j,i}(s+\bar{s}_k; s, \Delta s)}
$$
where $f_{s_k}^{j,i}(s+\bar{s}_k; s, \Delta s)$ denotes the real value, $\bar{f}_{s_k}^{j,i}(s+\bar{s}_k; s, \Delta s)$ denotes the value calculated according to the approximation formula, be less than any arbitrary percentage $\zeta$, we can require the error of $\bar{\Psi}_{ij}(s+\Delta s, {\bf u},\Delta\bar{s}): \epsilon$,  where $\Delta\bar{s}=\frac{-\ln(1-\epsilon)}{K\cdot\lambda_{\max}(s+\Delta s)}$, to satisfy the following conditions:
$$
\left\{
\begin{array}{lll}
&&2^{\frac{\bar{s}_k\cdot K\cdot\lambda_{max}(s+\Delta s)}{-\ln(1-\epsilon)}}\left[(1+\epsilon\cdot e^{\frac{-\ln(1-\epsilon)}{K}})^{\frac{\bar{s}_k\cdot K\cdot\lambda_{max}(s+\Delta s)}{-\ln(1-\epsilon)}+1}-1\right]<\frac{\zeta}{2}\\
&&2^{\frac{(\Delta s-\bar{s}_k)\cdot K\cdot\lambda_{max}(s+\Delta s)}{-\ln(1-\epsilon)}}\left[(1+\epsilon\cdot e^{\frac{-\ln(1-\epsilon)}{K}})^{\frac{(\Delta s-\bar{s}_k)\cdot K\cdot\lambda_{max}(s+\Delta s)}{-\ln(1-\epsilon)}+1}-1\right]<\frac{\zeta}{2}.
\end{array}
\right.
$$
\end{proposition}

\begin{proof}
Notice that when $s_1<s_2$, the following relationship
$$
\frac{-\ln(1-\epsilon)}{K\cdot\lambda_{\max}(s_2)}\le\frac{-\ln(1-\epsilon)}{K\cdot\lambda_{\max}(s_1)}
$$
would always be valid. That is to say, when we choose the numerical time step size $\Delta \bar{s}$ to ensure the error of $\bar{\Psi}_{ij}(s+\Delta s, {\bf u},\Delta\bar{s})$ be less than $\epsilon$, this step size would also ensure the error of $\bar{\Psi}_{ij}(s_0, {\bf u},\Delta\bar{s})$ where $s_0\in [0, s+\Delta s]$ be less than $\epsilon$ as well.
Because $P(X_t=x_j)$ and $\bar{\Psi}_{ij}(s_0, {\bf u},\Delta\bar{s})$ are always less than $1$,
from the expressions for calculating $f_{s_k}^{j,i}(s+\bar{s}_k; s, \Delta s)$ above,
to make sure that the error be less than $\zeta$, we have the following relationships
$$
\sum_{l_1=0,1}\cdots\sum_{l_{M_1}=0,1}\left(\frac{(\bar{\Psi}_{jl_1}+\epsilon)\cdot(\bar{\Psi}_{l_1l_2}+\epsilon)\cdots(\bar{\Psi}_{l_{M_1}l}+\epsilon)-\bar{\Psi}_{jl_1}\cdot\bar{\Psi}_{l_1l_2}\cdots\bar{\Psi}_{l_{M_1}l}}{\bar{\Psi}_{jl_1}\cdot\bar{\Psi}_{l_1l_2}\cdots\bar{\Psi}_{l_{M_1}l}}\right)<\frac{\zeta}{2}
$$
which implies
$$
\sum_{l_1=0,1}\cdots\sum_{l_{M_1}=0,1}\left((1+\frac{\epsilon}{\bar{\Psi}_{jl_1}})\cdot(1+\frac{\epsilon}{\bar{\Psi}_{l_1l_2}})\cdots(1+\frac{\epsilon}{\bar{\Psi}_{l_{M_1}l}})-1\right)<\frac{\zeta}{2}
$$
and
$$
\sum_{\bar{l}_1=0,1}\cdots\sum_{\bar{l}_{M_2}=0,1}\left(\frac{(\bar{\Psi}_{l\bar{l}_1}+\epsilon)\cdot(\bar{\Psi}_{\bar{l}_1\bar{l}_2}+\epsilon)\cdots(\bar{\Psi}_{\bar{l}_{M_2}i}+\epsilon)-\bar{\Psi}_{l\bar{l}_1}\cdot\bar{\Psi}_{\bar{l}_1\bar{l}_2}\cdots\bar{\Psi}_{\bar{l}_{M_2}i}}{\bar{\Psi}_{l\bar{l}_1}\cdot\bar{\Psi}_{\bar{l}_1\bar{l}_2}\cdots\bar{\Psi}_{\bar{l}_{M_2}i}}\right)<\frac{\zeta}{2}
$$
which implies
$$
\sum_{\bar{l}_1=0,1}\cdots\sum_{\bar{l}_{M_2}=0,1}\left((1+\frac{\epsilon}{\bar{\Psi}_{l\bar{l}_1}})\cdot(1+\frac{\epsilon}{\bar{\Psi}_{\bar{l}_1\bar{l}_2}})\cdots(1+\frac{\epsilon}{\bar{\Psi}_{\bar{l}_{M_2}i}})-1\right)<\frac{\zeta}{2}
$$
where $l=0,1$.
Notice that $\bar{\Psi}_{ij}(s_0, {\bf u},\Delta\bar{s}), s_0\in[s, s+\Delta s]$ in the above would always be greater than $e^{-\lambda_{max}(s+\Delta s)\cdot\Delta\bar{s}}$ which is equal to $e^{\frac{\ln(1-\epsilon)}{K}}$.
Thus we can replace each $\bar{\Psi}_{ij}(s_0, {\bf u},\Delta\bar{s})$ in the above with $e^{\frac{\ln(1-\epsilon)}{K}}$ to find $\epsilon$ according to $\zeta$.
Also, note that
$$
M_1\le\frac{\bar{s}_k\cdot K\cdot\lambda_{max}(s+\Delta s)}{-\ln(1-\epsilon)}\quad {\rm and } \quad
M_2\le\frac{(\Delta s-\bar{s}_k)\cdot K\cdot\lambda_{max}(s+\Delta s)}{-\ln(1-\epsilon)},
$$
then the above equations can be rewritten as follows:
$$
\left\{
\begin{array}{lll}
&&2^{M_1}\left[(1+\frac{\epsilon}{e^{\frac{\ln(1-\epsilon)}{K}}})^{M_1+1}-1\right]\le 2^{\frac{\bar{s}_k\cdot K\cdot\lambda_{max}(s+\Delta s)}{-\ln(1-\epsilon)}}\left[(1+\frac{\epsilon}{e^{\frac{\ln(1-\epsilon)}{K}}})^{\frac{\bar{s}_k\cdot K\cdot\lambda_{max}(s+\Delta s)}{-\ln(1-\epsilon)}+1}-1\right]<\frac{\zeta}{2}\\
&&2^{M_2}\left[(1+\frac{\epsilon}{e^{\frac{\ln(1-\epsilon)}{K}}})^{M_2+1}-1\right]\le 2^{\frac{(\Delta s-\bar{s}_k)\cdot K\cdot\lambda_{max}(s+\Delta s)}{-\ln(1-\epsilon)}}\left[(1+\frac{\epsilon}{e^{\frac{\ln(1-\epsilon)}{K}}})^{\frac{(\Delta s-\bar{s}_k)\cdot K\cdot\lambda_{max}(s+\Delta s)}{-\ln(1-\epsilon)}+1}-1\right]<\frac{\zeta}{2}
\end{array}
\right.
$$
\end{proof}

All the conditions related to the relative errors of $f_{s_k}^{j,i}(s+\bar{s}_k; s, \Delta s)$, $f_{t_k}^{j,i}(s+\bar{t}_k; \beta, s, \Delta s)$ and $P(X_t=x_j, {\rm no \ jump \ or \ default \ in} [0,t])$ similar to the above proposition should be satisfied to find a suitable $\epsilon$. Therefore, the relative errors are controlled and the error of $P(X_t=x_i|\mathcal{F}_t)$ is also controlled. \\
We remark that suppose the expiry time is denoted as $T_{expiry}$, then all $\lambda_{max}(s_0), s_0\in[0, T_{expiry}]$ in proposition $5$ and proposition $6$ could simply be replaced by $\lambda_{max}=\lambda_{max}(T_{expiry})$.

\section{Numerical Experiments}\label{Numerical Experiments}

In the following numerical experiments, for the configuration of the parameters value in the hidden Markov chain $X_t$, we let the transition rates be $\theta_0=0.1$ and $\theta_1=0.1$, the initial state $x_0=0$. For the observable chain $Y_t$, we set the transition rates
$$
\eta_0(x)=\left\{
\begin{array}{ll}
0.1, &x=x_0\\
0.2, &x=x_1
\end{array}
\right.
$$
and
$$
\eta_1(x)=\left\{
\begin{array}{ll}
0.2, &x=x_0\\
0.1, &x=x_1.
\end{array}
\right.
$$
and the initial state is $y_0=0$ as we assumed.
The risk-free interest rate $r$ is assumed to be $5\%$.

\subsection{Numerical Example 1}

We consider the pricing of Credit Default Swaps (CDS).
Assume that the buyer of the CDS agrees to pay premiums
to the seller continuously over time at a fixed rate until the expiration time of the CDS contract.
If the reference asset defaults prior to the expiry,
then the seller will pay $\$1$ to the buyer.
Denote the seller as entity A, buyer as entity B and the reference asset of the CDS as entity C.
Denote $\tau^A, \tau^B, \tau^C$ the default times and $\lambda_A, \lambda_B, \lambda_C$
the default intensities of entities A, B and C, respectively.
Here the default intensities of these homogeneous three entities are assumed in the following form:
$$
\lambda_i(t)= a + b\cdot X(t)+c\cdot \left(\sum_{j\ne i}1_{\{\tau^j\le t\}}\right),
\quad i, j = A,B,C
$$
where $a, b$ and $c$ are constants, $X(t)$ represents the hidden state process and
$\sum_{j\ne i}1_{\{\tau^j\le t\}}$ represents the default processes which are observable.
Let $y$ be the fixed premium rate, and suppose the issue time of the swap contract is $0$, the expiry time is $T$, and we are at time $s$, then the present value of the premium payment from the buyer should be
$$
E\left[ \int_{0}^{T}e^{- r s }y1_{\{s<\tau^A,s<\tau^B,s<\tau^C\}}ds \right].
$$
This means if any one of the three entities defaults,
the buyer of the CDS contract would stop paying the premium.
Similarly, the present the value of the seller should be
$$
E \left[ e^{-r T}1_{\{T<\tau^A,T<\tau^B,\tau^C\le T\}}ds \right].
$$
According to these two expressions,
one can obtain the premium of the CDS in the following form:
$$
y=\frac{E\left(e^{- r T}1_{\{T<\tau^A,T<\tau^B,\tau^C\le T\}}ds \right)}{E\left(\int_{0}^{T}e^{- r s}1_{\{s<\tau^A,s<\tau^B,s<\tau^C\}}ds\right)}.
$$
From the above formula, we know that to calculate $y$,
we need to compute the joint density function $f(s<\tau^A,s<\tau^B,s<\tau^C)$ and the joint probability  $P(T<\tau^A,T<\tau^B,\tau^C\le T)$.
Notice that $f(s<\tau^A,s<\tau^B,s<\tau^C)$ is actually equal to $f(\tau^1>s)$
where $\tau^1$ denotes the time of the first default out of the $3$ entities,
and $P(T<\tau^A,T<\tau^B,\tau^C\le T)=P(\tau^1\le T<\tau^2)$.
Here $\tau^1$ has the same meaning as before, $\tau^2$ denotes the time of the second default in the reference portfolio.
Then we can apply the methods introduced in the previous sections to calculate the fixed premium rate $y$.
The base setting of parameters are as follows.
For the contagion factors, we let $a=1, b=0.1, c=0.1$.
The expiry $T$ is $5$ years, and the initial time is $0$.
We change the coefficients $a, b$ and $c$ in the expressions of default intensities separately,
and each time we keep the remaining coefficients unchanged to investigate the change in the CDS premium rate $y$.

\begin{figure}[htbp]
\centering
\subfigure[premium $y$ change with respect to $a$]{
\begin{minipage}{0.36\textwidth}
\centering
\includegraphics[width=2.2in]{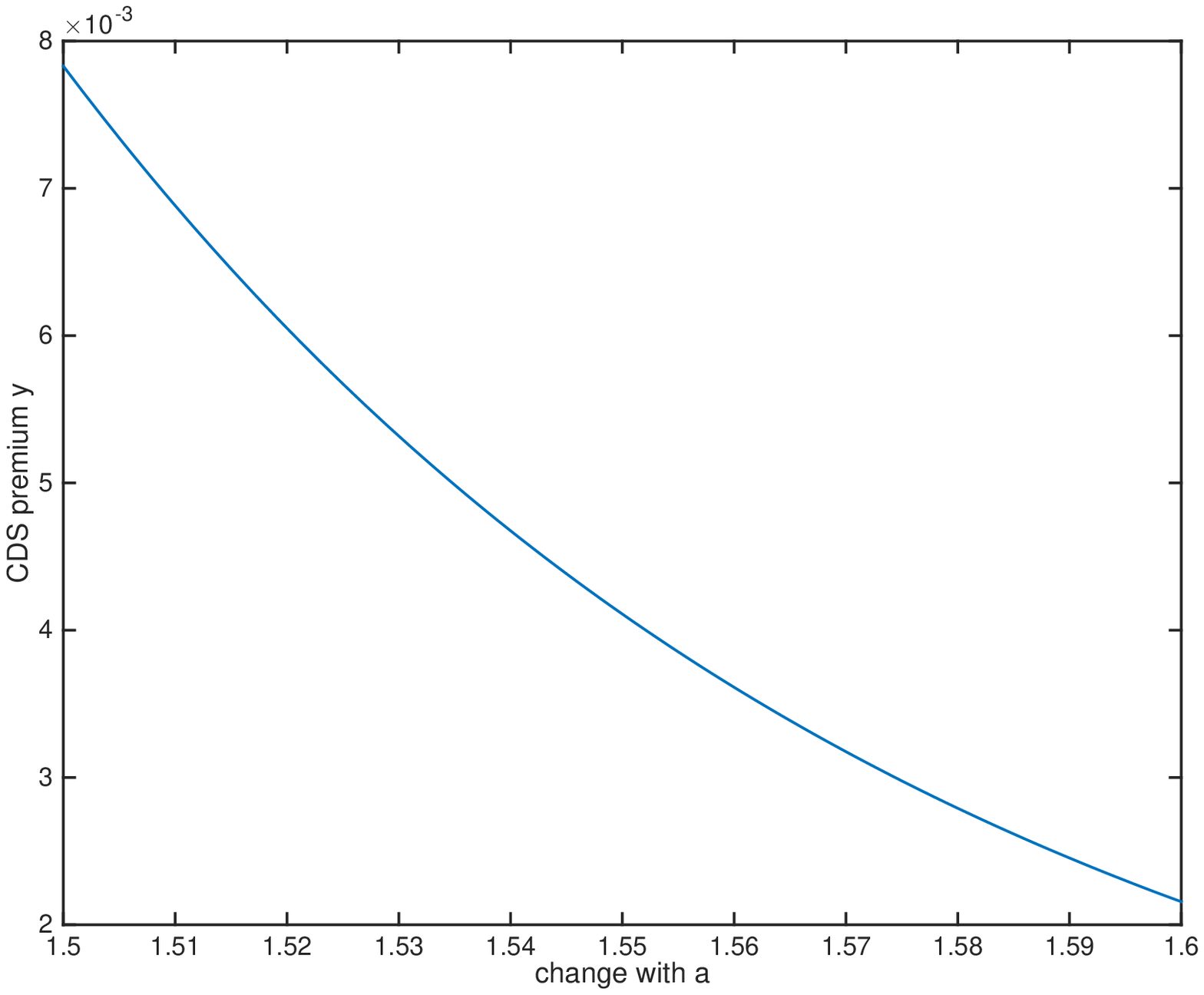}
\end{minipage}%
}%
\subfigure[premium $y$ change with respect to $b$]{
\begin{minipage}{0.36\textwidth}
\centering
\includegraphics[width=2.2in]{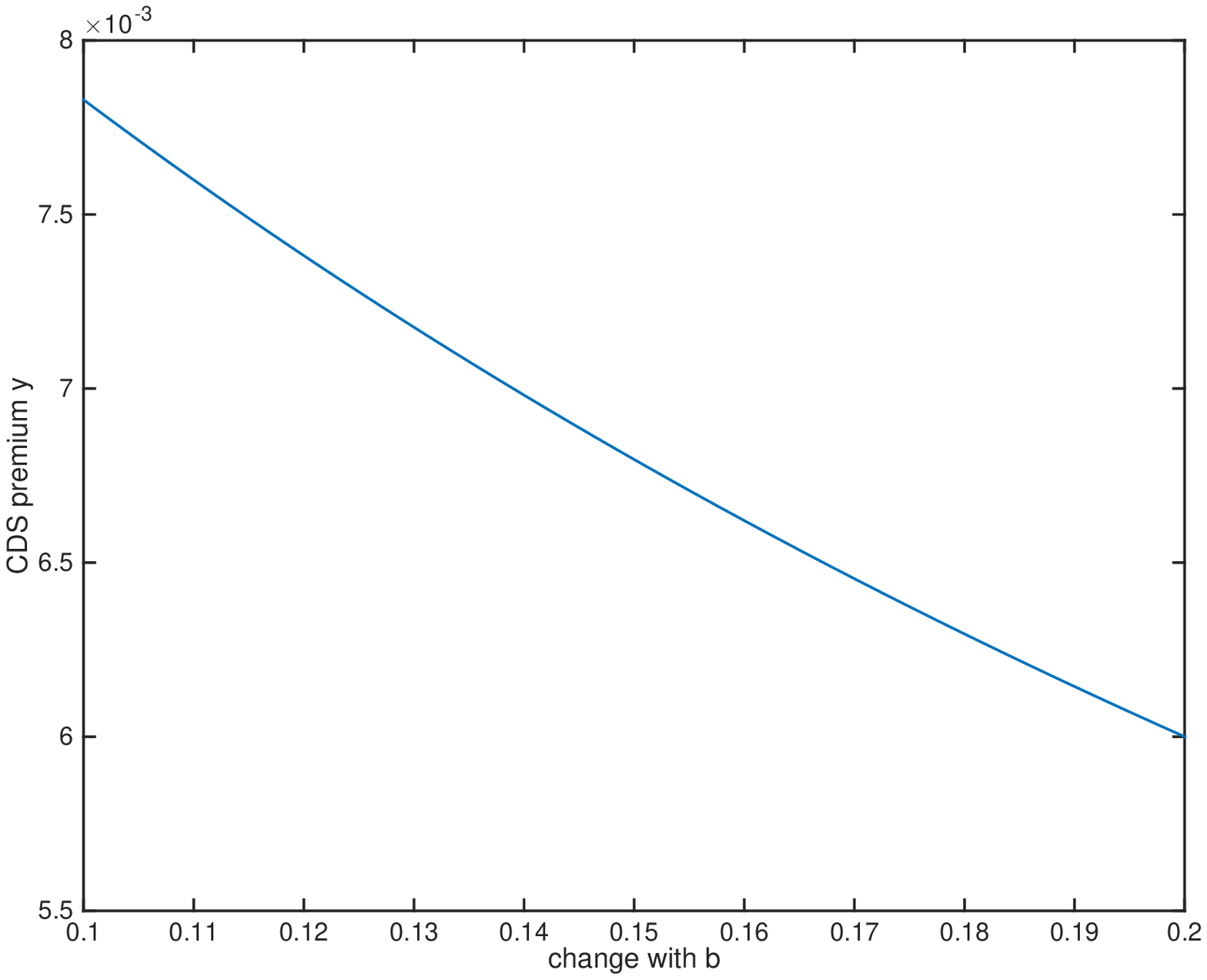}
\end{minipage}%
}%
\subfigure[premium $y$ change with respect to $c$]{
\begin{minipage}{0.36\textwidth}
\centering
\includegraphics[width=2.2in]{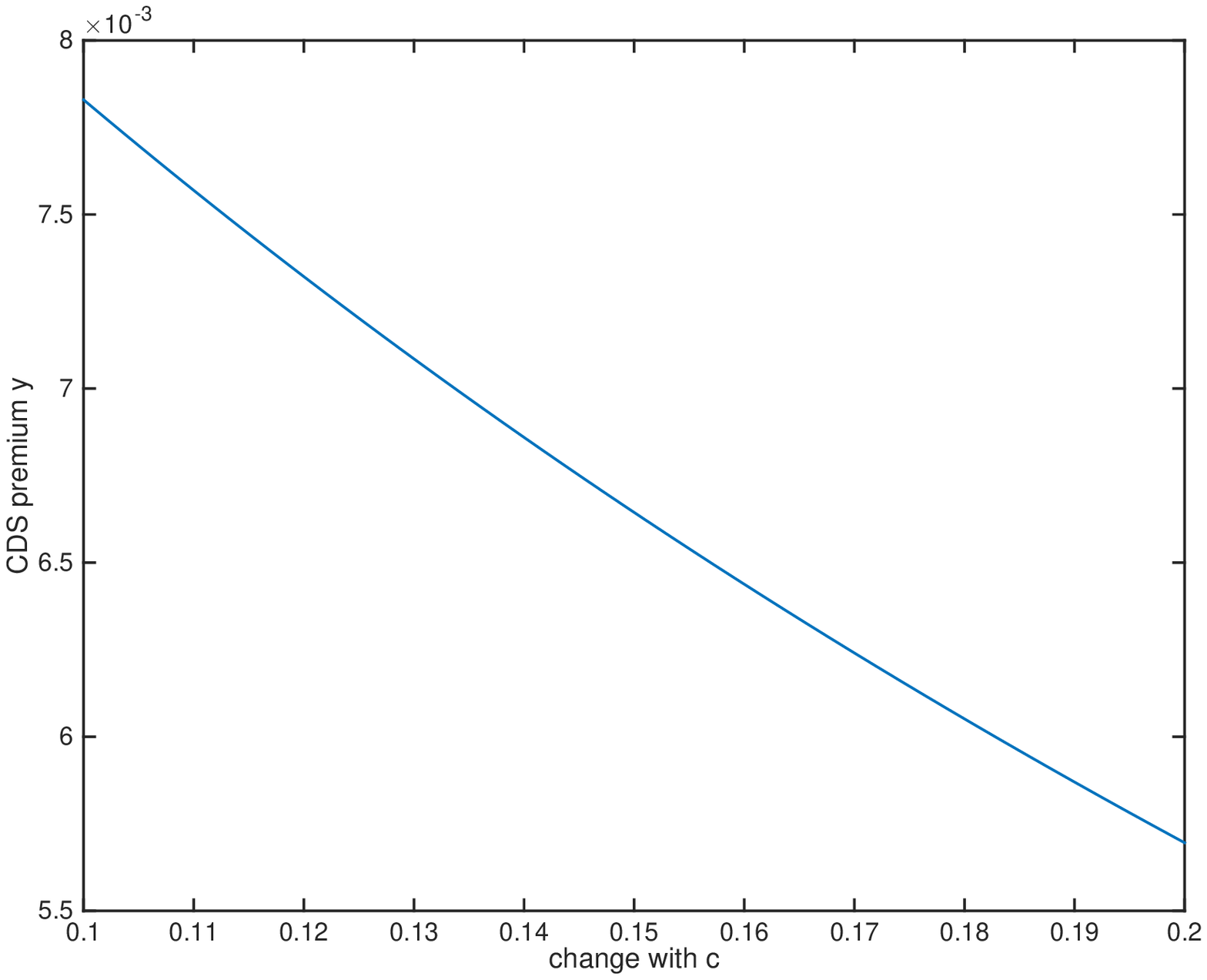}
\end{minipage}
}
\caption{Change of premium $y$ with coefficients}
\end{figure}



From the above three figures, we find that the value of CDS premium rate $y$
decreases as the coefficients $a, b$, and $c$ increase.

\subsection{Numerical Example 2}

We then consider a $k$th-to-default basket CDS contact.
Assume that our portfolio contains $K=10$ homogeneous entities,
if $k$ entities out of this portfolio default prior to the expiry time, then $\$1$ will be paid.
For simplicity, this payment only occurs at the expiry time, but the payment of premium occurs at the initial time.
Similar to the previous experiment, the entity $i$'s default intensity is given by
$$
\lambda_i(t)=a+b\cdot X(t)+c\cdot \left(\sum_{j\ne i}1_{\{\tau^j\le t\}}\right), \quad i,j=1,2,\cdots,K.
$$
The value of this $k$th-to-default basket CDS at time $t$ can be written as
$$
V_k(t)=\exp\{-r(T-t)\}P(\tau^k\le T\mid \mathcal{F}_t)
$$
where $\tau^k$ denotes the $k$th-to-default time.
For the state of chain $X$, $x_0$ and $x_1$ represent the ``good'' and ``bad'' economic state, respectively.
While States $y_0$ and $y_1$ of chain $Y$ represent the delayed information of ``bad'' economic state and ``good'' economic state, respectively.
Here we also assume that the total number of entities in the portfolio is $K=10$.
The calculation of $P(\tau^k\le T\mid \mathcal{F}_s)$ can be obtained
from $1-P(\tau^{j_k}>T\mid \mathcal{F}_s)$ where
$$
P(\tau^{j_k}>t\mid \mathcal{F}_s)=\sum_{i=m}^{k-1}P(\tau^{j_i}\le t<\tau^{j_{i+1}}\mid \mathcal{F}_s).
$$
The calculation of the probability $P(\tau^{j_i}\le t<\tau^{j_{i+1}}\mid \mathcal{F}_s)$ is similar to the calculation of $P(\tau^1\le t<\tau^2)$ in Experiment $1$.
Without loss of generality, for simplicity, we consider the $1$st-to-default basket CDS as $k=1$.
We further assume that the initial time is $0$, and that we are at time $t=10$ days now, and that the expiry time is $T=100$ days.
In the following experiments, we consider two scenarios.
In Scenario $1$, there is no jump in chain $Y$ and default observed by expiry time.
In Scenario $2$, there is one jump in chain $Y$ between day $21$ and day $22$ but no default observed by expiration.
According to the assumptions presented in Section $2$, we know that the initial state of chain $X$ is $x_0=0$ and the initial state of chain $Y$ is $y_0=0$.
In addition, let the coefficients in default intensities be $a=0.001, b=0.001$, and $c=0.001$.
Then one can see the change of basket CDS values from day to day,
and here we only provide the values from day $t=10$ to day $t=50$ as an example.

\begin{figure}
\centering
\includegraphics[width=6.0in]{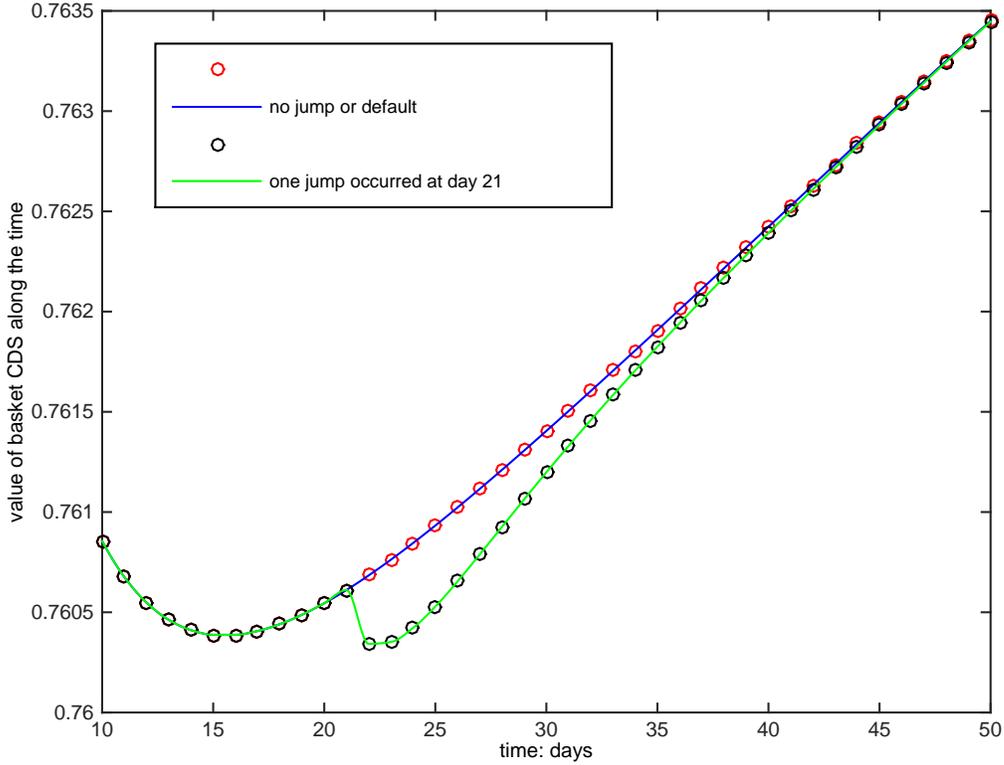}
\caption{Change of CDS's value from day to day with the 1st default intensities}
\end{figure}

From the figure we can see that as time goes by,
the general tendency of basket CDS's value is increasing.
When there is one jump in chain $Y$ from state $y_0$ to state $y_1$,
the value will drop suddenly.
It is because at the beginning, the information of chain $Y$ reflected a ``bad'' economic condition,
when it changed to state $y_1$ which representing a ``good'' economic state,
intuitively, the probability of defaults will drop suddenly,
and the value of basket CDS will therefore drop suddenly as well.

As we mentioned before, our model and methods may be applicable to various forms of default intensities. Therefore, we further consider another form of default intensities which decay exponentially with time. The expression is as follows:
$$
\lambda_i(t)=\left(a+c\cdot\sum_{j\ne i}1_{\{\tau^j\le t\}}\right)e^{-t}+b\cdot X(t), \quad i,j=1,2,\ldots,K.
$$
Same as before, all parameters in this default intensity keep the same as the previous case,
then we can also calculate the value of basket CDS and observe it from day to day.

\begin{figure}
\centering
\includegraphics[width=6.0in]{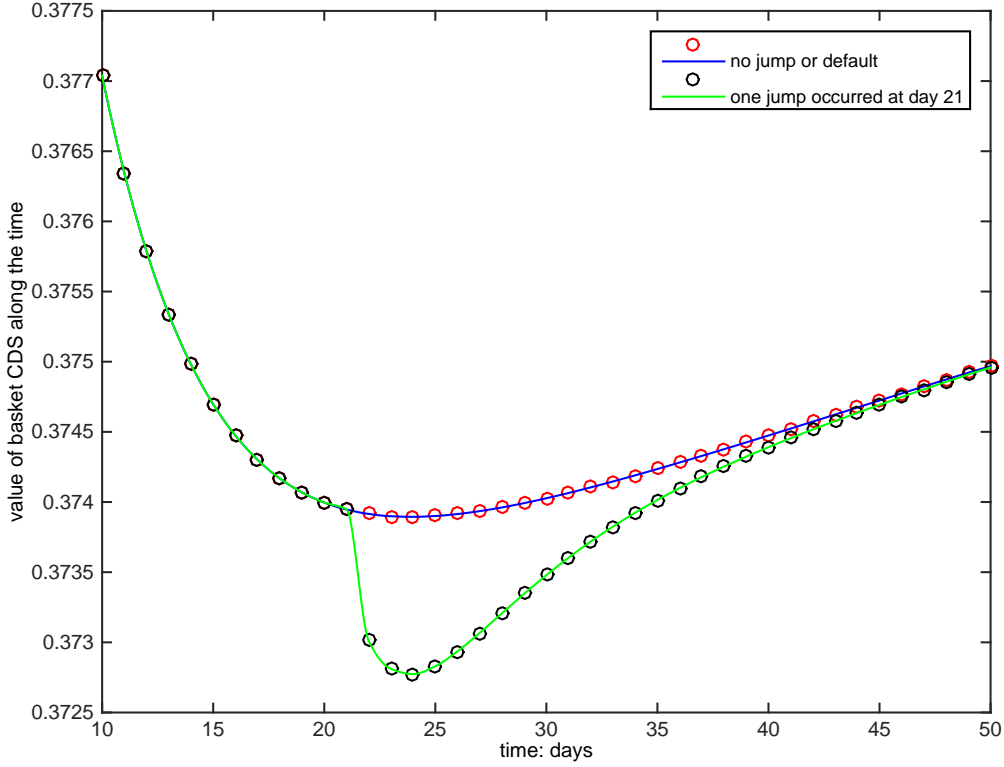}
\caption{Change of CDS's value from day to day with the 2nd default intensities}
\end{figure}

From the figure, we notice that the overall value based on this form of default intensity
is smaller than the previous one.
The phenomenon can be explained as follows.
As the default intensity exponentially decreased with time,
the default probability will become smaller accordingly and therefore the value of basket CDS.
For the same reason and similar explanations like before, the value will also jump down suddenly when one jump in observable chain $Y$ from state $y_0$ to $y_1$ occurred.

\section{Concluding Remarks}\label{Concluding Remarks}

In this paper we present a reduced-form intensity-based credit risk model with a
hidden Markov process modeling the evolution of economic condition over time.
We also discuss a method to extract the underlying hidden state process from observable processes:
the default processes and the stochastic process which reflects
the delayed and noisy information about the hidden state process. The method may have a wide range of applications.
Based on this, we develop a closed-form expression
to obtain the joint default distribution with the hidden state process. After deriving this general formula, 
for the homogeneous contagion portfolio, we also give analytical formulas
for the distribution of ordered default times. Beside, we extend the total hazard construction method to get the joint distribution of default times for hidden markov models.
We remark that the methods discussed may be applicable to various forms of default intensities.
Algorithms for practical implementation of the methods are presented and their uses for pricing credit derivatives are illustrated.
In the numerical experiments, we consider valuations for the CDSs premium rates of the regular and basket type
with different expressions of default intensities which cover an exponential decay and a stochastic intensity process.
We also study the sensitivities of premium rates with respect to changes in the underlying parameters in the regular CDS as an example.

\section*{Acknowledgements}

This research work was supported by
Research Grants Council of Hong Kong under Grant
Number 17301214 and HKU CERG Grants
and HKU strategic Research Theme in Information and Computing.


\begin{thebibliography}{1}

\bibitem{BS1973}
F. Black and M. Scholes,
{\em The pricing of options and corporate liabilities},
Journal of Political Economy, 81(3), 637--654, 1973

\bibitem{BPT2006}
D. Brigo, A. Pallavicini \& R. Torresetti,
{\em Calibration of CDO tranches with the dynamical generalized-Poisson loss
model}, Working Paper, Banca IMI, 2006,
available at  {\it http://papers.ssrn.com/sol3/papers.cfm?abstract\_id=900549}.

\bibitem{CM2011}
R. Cont and A. Minca,
{\em Recovering portfolio default intensities implied by CDO quotes},
Mathematical Finance. doi: 10.1111/j.1467-9965.2011.00491.x, 2011

\bibitem{DL2001}
M. Davis and V. Lo,
{\em Modeling default correlation in bond portfolios},
in C. Alexander (Ed.),
Mastering Risk Volume 2: Applications, Prentice Hall, 141-151, 2001.


\bibitem{EJY}
R. J. Elliott, M. Jeanblanc and M. Yor, (2000), {\em On Models of Default Risk},
Mathematical Finance, 10(2), 179–195.

\bibitem{ES} R.J. Elliott and T.K. Siu, (2013), {\em An HMM Intensity-Based Credit Risk Model
and Filtering}, State-Space Models and Applications in Economics and Finance, Statistics and Econometrics in Finance, Volume 1, (edited by Yong Zeng and Shu Wu), Springer-Verlag, pp. 169-184.

\bibitem{ESF} R.J. Elliott, T.K. Siu and E.S. Fung, (2014),
 {\em A Double HMM Approach to
Altman Z-scores and Credit Ratings}, Expert Systems With Applications,
41(4-2), pp. 1553-1560.

\bibitem{FR2010}
 Frey, R., and W. Runggaldier, (2010), {\em Pricing credit derivatives under incomplete information: a nonlinear-filtering approach}, Finance and Stochastics, 14 (4) pp. 495-526.

\bibitem{FR2011}
 Frey, R., and W. Runggaldier, (2011), {\em Nonlinear Filtering in Models for Interest Rate and Credit Risk}, Chapter 32 in Handbook of Nonlinear Filtering, D. Crisan, B. Rozovski,
eds., Oxford University Press.


\bibitem{FS}
Frey, R., and T. Schmidt,
(2011) Filtering and Incomplete Information in Credit Risk, Chapter 7 in Recent Advancements in the Theory and Practice of Credit Derivatives,
Damiano Brigo, Tom Bielecki and Frederic Patras, ed., Wiley, New Jersey.






\bibitem{G2008}
Giesecke, K. (2008),
 {\em Portfolio Credit Risk: Top-Down vs. Bottom-Up Approaches},
  in: Frontiers in Quantitative Finance: Credit Risk and Volatility Modeling, R. Cont (Ed.), Wiley.

\bibitem{GG2004}
K. Giesecke and L. Goldberg,
{\em Sequential defaults and incomplete information},
Journal of Risk, 7(1), 1--26, 2004.

\bibitem{GGD2005}
K. Giesecke, L. Goldberg and X. Ding,
{\em A top down approach to multi-name credit},
Operations Research, 59(2), 283--300, 2011.

\bibitem{Gu1}
J. Gu, W. Ching, T. Siu and H. Zheng,
\emph{ On pricing basket credit default swaps},
Quantitative Finance, 13(12), 1845--1854, 2013.

\bibitem{Gu2}
J. Gu, W. Ching and H. Zheng,
\emph{ A hidden Markov reduced-form risk model},
Computational Intelligence for Financial Engineering \& Economics, 2014 IEEE Conference, 190--196, 2014.

\bibitem{DG2001}
D. Duffie and N. Garleanu,
{\em Risk and valuation of collateralized debt obligations},
Financial Analysts Journal, 57(1), 41--59, 2001.

\bibitem{DSW2006}
D. Duffie, L. Saita and K. Wang,
{\em Multi-period corporate default prediction with stochastic covariates},
Journal of Financial Economics, 83(3), 635--665, 2006.

\bibitem{JT1995}
R. Jarrow and S. Turnbull,
{\em Pricing derivatives on financial securities subject to credit risk},
Journal of Finance, 50, 53--86, 1995.

\bibitem{JY2001}
R. Jarrow and F. Yu,
{\em Counterparty risk and the pricing of defaultable securities},
Journal of Finance, 56(5), 555--576, 2001.

\bibitem{David}
D. Lando,
\emph{On Cox processes and credit risky securities},
Review of Derivatives Research, 2,  99--120, 1998.

\bibitem{LR2008}
F. Longstaff and A. Rajan,
{\em An empirical analysis of collateralized debt obligations},
Journal of Finance, 63(2), 529--563, 2008.

\bibitem{Merton}
R. Merton,
{\em On the pricing of corporate debt: the risk structure of interest rates},
Journal of Finance, 29(2), 449--470, 1974.

\bibitem{MU1998}
D. Madan and H. Unal,
{\em Pricing the risks of default},
Review of Derivatives Research, 2(2-3), 121--160, 1998.

\bibitem{Norros}
I. Norros,
\emph{ A compensator representation of multivariate life length distributions, with applications},  Scand. J. Stat., 13, 99--112, 1986.

\bibitem{Shaked}
M. Shaked  and G. Shanthikumar,
\emph{ The multivariate hazard construction},
Stoch. Proc. Appl, 24,  241--258, 1987.

\bibitem{SS2001}
P. Sch\"onbucher and D. Schubert,
{\em Copula-dependent default risk in intensity models},
Working paper, Universit at Bonn, 2001.

\bibitem{Yu}
F. Yu,
\emph{Correlated defaults in intensity-based models},
 Mathematical Finance, 17(2), 155--173, 2007.

\bibitem{Zheng}
H. Zheng  and  L. Jiang,
\emph{ Basket CDS pricing with interacting intensities},
Finance and stochastics, 13, 445--469, 2009.

\end{thebibliography}
\end{document}